\documentclass[epj]{svjour}
\usepackage{graphicx,color,amssymb}
\usepackage[numbers,sort&compress,merge]{natbib}

\begin{document}
\pagestyle{plain}
\title{Implications of LHC Searches on SUSY Particle Spectra}
\subtitle{The pMSSM Parameter Space with Neutralino Dark Matter}
\author{A. Arbey\inst{1}\inst{2}\inst{3} \and 
        M. Battaglia\inst{2} \inst{4} \inst{5} \and
        F. Mahmoudi\inst{2} \inst{6}
}                     
%
%

\institute{Universit\'e de Lyon, France; Universit\'e Lyon 1, CNRS/IN2P3, UMR5822 IPNL, 
F-69622~Villeurbanne Cedex, France \and
CERN, CH-1211 Geneva 23, Switzerland \and
Centre de Recherche Astrophysique de Lyon, Observatoire de Lyon, Saint-Genis Laval Cedex, F-69561, France; 
CNRS, UMR 5574; Ecole Normale Sup\'erieure de Lyon, Lyon, France \and
Santa Cruz Institute of Particle Physics, University of California, Santa Cruz,
CA 95064, USA \and
Lawrence Berkeley National Laboratory, Berkeley, CA 94720, USA \and
Clermont Universit\'e, Universit\'e Blaise Pascal, CNRS/IN2P3, LPC, BP 10448, 63000 Clermont-Ferrand, France}
\date{}
%
\abstract{
We study the implications of LHC searches on SUSY particle spectra using flat scans of the 19-parameter pMSSM phase space. We apply constraints from flavour physics, $g_\mu-2$, dark matter and earlier LEP and Tevatron searches. The sensitivity of the LHC SUSY searches with jets, leptons and missing energy is assessed by reproducing with fast simulation the recent CMS analyses after validation on benchmark points. We present results in terms of the fraction of pMSSM points compatible with all the 
constraints which are excluded by the LHC searches with 1~fb$^{-1}$ and 15~fb$^{-1}$ as a function of the mass of strongly and weakly interacting SUSY particles. We also discuss the suppression of Higgs production cross sections for the MSSM points not excluded and contrast the region of parameter space tested by the LHC data with the constraints from dark matter direct detection experiments.
} 
\maketitle
\section{Introduction}
\label{sec:1}
Supersymmetry (SUSY) has emerged over the past two decades as possibly the best motivated model of new physics beyond 
the Standard Model (SM). Together with stabilising the masses at the electroweak scale, it provides gauge coupling 
unification and viable candidates for cold dark matter. Expectations for an observation of supersymmetric partners of 
the strongly interacting SM particles in the early stage of the LHC run have been high. In fact, global fits to constrained 
SUSY models, such as the CMSSM, including data from flavour physics, lower energy experiments and relic dark matter density  
have favoured SUSY particle masses below, or around, 1~TeV \cite{Allanach:2007qk,Allanach:2008tu,Buchmueller:2008qe,Trotta:2008bp}. Recent searches for supersymmetric particles 
by the ATLAS and CMS experiments at the LHC have now excluded most of this portion of the parameter space for these models, 
raising questions about the range of SUSY particle masses still allowed by the present data. Several studies have been carried 
out to evaluate the impact of LHC and other data on SUSY parameters. Most of these studies considered the highly constrained 
models with a small number of free parameters and large correlations between the masses of supersymmetric particles, 
which had been used for earlier benchmark studies and model parameter fits \cite{Buchmueller:2011ki,Bertone:2011nj}. In these constrained models, the LHC 
searches have resulted in a significant exclusion of masses at, or beyond, 1~TeV for the majority of SUSY particles. 
However, these models are not representative of a generic minimal supersymmetric extension of the Standard Model (MSSM), 
where the particle mass parameters are independent. The phenomenological MSSM (pMSSM), with its 19-parameter phase 
space~\cite{Djouadi:2002ze}, was proposed to reduce the theoretical prejudices of these constraints. Two pioneering studies 
considered the pMSSM for a global Bayesian fit using low energy and dark matter constraints to make inferences on SUSY 
particle masses~\cite{AbdusSalam:2009qd} and the sensitivity of LHC searches~\cite{Berger:2008cq,Conley:2010du,Conley:2011nn}. More recently, two further studies addressed the sensitivity of LHC to pMSSM scenarios based on ATLAS \cite{Allanach:2011ej} and CMS \cite{Sekmen:2011cz} analysis strategies.

In this paper we report a study of the impact of the present LHC data at 7~TeV in the pMSSM with neutralino dark matter. Compared to the previous pMSSM study of Ref.~\cite{Conley:2011nn}, we use here a larger scan statistics of almost 25M points associated to a broad range of the parameters, in particular for the gluino mass, in a flat scan, we adopt the selection criteria and results of the latest analyses on 1~fb$^{-1}$ of LHC data and the updated constraints from rare $B$ decays, $g_{\mu}-2$ and $\Omega_{\mathrm{DM}} h^2$ and present our results in terms of fractions of pMSSM points excluded by the LHC data for 1~fb$^{-1}$ and a 15~fb$^{-1}$ projection without adopting a likelihood weighting. The main purposes of this study are to find which information on non-constrained SUSY scenarios are implied by LHC searches, and to investigate whether there are parameter regions not probed within the broad bounds on squark and gluino masses, possibly corresponding to distinct mass or coupling patterns. In addition, we aim to understand to which extent the spectrum of weakly 
interacting SUSY particles, which are not directly probed by the LHC searches, is pushed towards higher masses as a 
consequence of the LHC results. 

This paper presents the scan and analysis method with results in the most studied neutralino 
dark matter scenario. To achieve our objectives, we perform flat scans over the parameter space of the pMSSM in order to probe different regions, without being interested in determining the best fit points of the model. Within these regions, we contrast the LHC sensitivity with results on dark matter direct detection and we comment on the suppression of event yields in $gg \rightarrow h^0 \rightarrow \gamma \gamma$ channel which is most relevant for the search for light Higgs boson at the LHC.

This work will be followed by analyses of the scan data in the framework of other scenarios, including 
light neutralino dark matter, gravitino dark matter, the NMSSM and in relation to the Higgs sector.

This paper is organised as follows. Section~\ref{sec:2} presents the scan of the SUSY parameters discussing 
the technical tools and the constraints from lower energy data and dark matter relic density adopted for our
analysis. The constraints from the LHC data are implemented by generating event samples for each SUSY point and 
applying the analyses performed by the CMS collaboration to these events after fast simulation. Section~\ref{sec:3} 
discusses the implementation of these analyses in our scan framework and their validation using the official CMS 
simulation and reconstruction.
In section~\ref{sec:4} we discuss the scan results in terms of both the sensitivity of the LHC searches with 1~fb$^{-1}$ and 15~fb$^{-1}$ to strongly- and weakly-interacting sparticles. 
These results are relevant to the assessment of the physics potential of the 7~TeV LHC operation and also to the planning of a future lepton collider, which should operate at a centre-of-mass energy above the gaugino 
and slepton pair production threshold to perform detailed studies of the spectroscopy of the weakly interacting 
SUSY particles, if SUSY is indeed realised in nature. Finally we discuss the spin independent $\tilde \chi p$ scattering cross section relevant to dark matter direct detection experiments, the suppression of light Higgs production and decay yields compared to the Standard Model and the fine tuning for the points not excluded by the LHC data. 

\section{pMSSM Scans}
\label{sec:2}
\begin{table}
\begin{center}
\begin{tabular}{|c|c|}
\hline
~~~~Parameter~~~~ & ~~~~~~~~~~Range~~~~~~~~~~\\
\hline\hline
$\tan\beta$ & [1, 60]\\
\hline
$M_A$ & [50, 2000]\\
\hline
$M_1$ & [-2500, 2500]\\
\hline
$M_2$ & [-2500, 2500]\\
\hline
$M_3$ & [50, 2500]\\
\hline
$A_d=A_s=A_b$ & [-2000, 2000]\\
\hline
$A_u=A_c=A_t$ & [-2000, 2000]\\
\hline
$A_e=A_\mu=A_\tau$ & [-2000, 2000]\\
\hline
$\mu$ & [-1000, 2000]\\
\hline
$M_{\tilde{e}_L}=M_{\tilde{\mu}_L}$ & [50, 2500]\\
\hline
$M_{\tilde{e}_R}=M_{\tilde{\mu}_R}$ & [50, 2500]\\
\hline
$M_{\tilde{\tau}_L}$ & [50, 2500]\\
\hline
$M_{\tilde{\tau}_R}$ & [50, 2500]\\
\hline
$M_{\tilde{q}_{1L}}=M_{\tilde{q}_{2L}}$ & [50, 2500]\\
\hline
$M_{\tilde{q}_{3L}}$ & [50, 2500]\\
\hline
$M_{\tilde{u}_R}=M_{\tilde{c}_R}$ & [50, 2500]\\
\hline
$M_{\tilde{t}_R}$ & [50, 2500]\\
\hline
$M_{\tilde{d}_R}=M_{\tilde{s}_R}$ & [50, 2500]\\
\hline
$M_{\tilde{b}_R}$ & [50, 2500]\\
\hline
\end{tabular}
 \end{center}
\caption{SUSY parameter ranges (in GeV when applicable).\label{tab:paramSUSY}}
\end{table}
In the pMSSM, all the soft SUSY breaking parameters are assumed to be real, in order to avoid new source of CP violation. The sfermion mass matrices and the trilinear coupling matrices are taken to be diagonal, to suppress FCNCs at tree level. In addition, the first and second sfermion generations are universal at low energy and the trilinear couplings are set to be equal for the three generations. With these assumptions, we are left with 19 free input parameters at the weak scale.

To explore the pMSSM parameter space, we perform flat scans by randomly varying the 19 parameters within the intervals given in 
Table~\ref{tab:paramSUSY}, with the SM parameters of Table~\ref{tab:paramSM}. Our scan strategy, which is not guided through the parameter space as in the case of Markov Chain Monte Carlo methods, may miss local features. This does not change in a significant way the fraction of excluded points and the other results given in Section~\ref{sec:4} as long as these regions are narrow. On the other hand our approach ensures a rather uniform sampling of a very broad parameter space, without the risk of missing sizeable but largely disconnected acceptable regions of parameters. In section \ref{sec:4} we discuss our scan coverage in terms of SUSY spectral features and particle properties. We also study how the fraction of valid MSSM points excluded by the LHC analysis changes with the SUSY parameter ranges adopted for the scan. We generate a total of 24.57 million points. 
\begin{table}
\begin{center}
\begin{tabular}{|c|c|}
\hline
~~~~Parameter~~~~ & ~~~~~~~~~~Value~~~~~~~~~~\\
\hline\hline
$\alpha_s(M_Z)$ & 0.1184\\
\hline
$\bar{m}_b(\bar{m}_b)$ & 4.19 GeV\\
\hline
$m_t^{\mbox{pole}}$ & 172.9 GeV\\
\hline
\end{tabular}
 \end{center}
\caption{SM parameter values \cite{Nakamura:2010zzi}.\label{tab:paramSM}}
\end{table}
In this study, it is imposed that the lightest supersymmetric particle is the neutralino. 

For the flavour related quantities, we adopt the numerical inputs of \cite{Mahmoudi:2007vz,*Mahmoudi:2008tp}.

\subsection{Simulation Tools}
\label{sec:2-1}

The study of the LHC sensitivity in the pMSSM parameter space requires the use of several programs from the spectrum generation to the computation 
of flavour and other low energy and dark matter related observables, as well as the simulation and analysis of events in 7~TeV $pp$ collisions. Throughout the analysis we use the  
SUSY Les Houches Accord (SLHA) format~\cite{Skands:2003cj,*Allanach:2008qq} to store the SUSY parameters, physical spectra and couplings, 
the HEPEVT format for simulated events and the {\tt ROOT} format \cite{Brun:1997pa} for results of the scans, the physics objects after fast simulation and the analysis results.

SUSY spectra are generated with {\tt SOFTSUSY 3.1.7}~\cite{Allanach:2001kg}. Results have been cross checked with {\tt SUSpect}~\cite{Djouadi:2002ze} 
and found to be compatible with the exception of few regions of the parameter phase space. The masses, widths 
and decay channels of Higgs bosons are obtained using {\tt FeynHiggs 2.8.4} \cite{Heinemeyer:1998yj,Heinemeyer:1998np} and {\tt HDECAY 3.53} \cite{Djouadi:1997yw}.
The widths and branching fractions of the SUSY particles are computed using {\tt SDECAY 1.3} \cite{Muhlleitner:2003vg}.
Flavour observables, the muon anomalous magnetic moment and the dark matter relic density are calculated with 
{\tt SuperIso Relic v3.1}~\cite{Mahmoudi:2007vz,*Mahmoudi:2008tp,Arbey:2009gu}. Also, the neutralino-nucleon scattering cross-sections which are needed to evaluate dark matter direct detection constraint are computed with {\tt micrOMEGAs 2.4} \cite{Belanger:2008sj}. 
Finally, we confront the spectrum of each pMSSM point to the exclusion bounds obtained from Higgs at LEP, Tevatron and LHC using 
{\tt HiggsBounds 3.4.0}~\cite{Bechtle:2008jh,Bechtle:2011sb}. 

We use the {\tt SuperIso} program as the central control program, which provides interfaces with the other above-mentioned programs through SLHA files. For each point giving a valid spectrum with $\tilde \chi^0_1$ LSP, we store the pMSSM parameters, 
SUSY particle spectra, branching fractions, and flavour and low energy observables in {\tt ROOT} format. We then impose the constraints on these observables and on the masses 
of the supersymmetric particles imposing the exclusion bounds from SUSY searches at LEP and the Tevatron and generate an SLHA file for each accepted point.  
These SLHA files, extended to include SUSY branching fractions, are used as input to {\tt PYTHIA 6.4.24}~\cite{Sjostrand:2006za} for event generation 
of inclusive SUSY production in $pp$ interactions at 7~TeV, using CTEQ5L parton distribution functions \cite{Lai:1999wy}. Cross sections are rescaled to their NLO values with 
the k-factors computed for each point by {\tt Prospino 2.0}~\cite{Beenakker:1996ed}. Generated events are passed through fast detector simulation 
using {\tt Delphes 1.9}~\cite{Ovyn:2009tx} tuned for the CMS detector. The output of the {\tt Delphes} simulation is saved in {\tt ROOT} format and 
followed by a custom implementation of the CMS SUSY analyses. Results are stored in a database which is subsequently read-in in {\tt ROOT} and used to 
determine the observability of each of the pMSSM scan points.

\subsection{Constraints}
\label{sec:2-2}
We apply constraints from flavour physics, the anomalous muon magnetic moment, relic dark matter and SUSY searches at LEP and the Tevatron.  
First we consider the flavour observables. The branching ratio of $ B \to X_s \gamma$ has been thoroughly studied for its high sensitivity 
to new physics effects (see for example \cite{Ellis:2007fu,Mahmoudi:2007gd,Heinemeyer:2008fb,Eriksson:2008cx}), since the SM contributions only appear at loop level. We follow the NNLO calculation of \cite{Misiak:2006zs,Misiak:2006ab} for the theoretical predictions. 
The theoretical uncertainties here are well under control and the experimental accuracy is good thanks to the $B$ factory data. This branching fraction provides 
strong constraints on the SUSY parameter space, especially for large $\tan \beta$, where it receives large enhancements. We use the following 95\% C.L. interval \cite{Mahmoudi:2008tp}:
\begin{equation}
 2.16 \times 10^{-4} < \mbox{BR}(B \to X_s \gamma) < 4.93 \times 10^{-4} \;.
\label{eq:bsg}
\end{equation}

The decay $B_s \to \mu^+ \mu^-$ is also a loop mediated process, which can receive extremely large SUSY contributions at large $\tan\beta$, and its branching fraction can be 
enhanced by several orders of magnitude compared to its SM prediction \cite{Choudhury:1998ze,Babu:1999hn,Ellis:2005sc,Carena:2006ai,Akeroyd:2011kd}. We use the two loop calculations based on \cite{Bobeth:2001sq,Bobeth:2001jm,Buras:2002vd,Bobeth:2004jz}. This decay mode has not yet been observed, but recently LHCb and CMS provided upper 
limits on its branching fraction which are only 3.4 times above the SM value~\cite{CMS_plus_LHCb}. The CMS and LHCb combination yields a 95\% C.L. upper limit of $1.1 \times 10^{-8}$. Here we add theoretical uncertainties and use a constraint of~\cite{Akeroyd:2011kd}:
\begin{equation}
\mbox{BR}(B_s \to \mu^+ \mu^-) < 1.26 \times 10^{-8} \;.
\label{eq:bsmumu}
\end{equation}

\begin{figure}[t!]
\begin{center}
\begin{tabular}{c}
\includegraphics[width=0.35\textwidth]{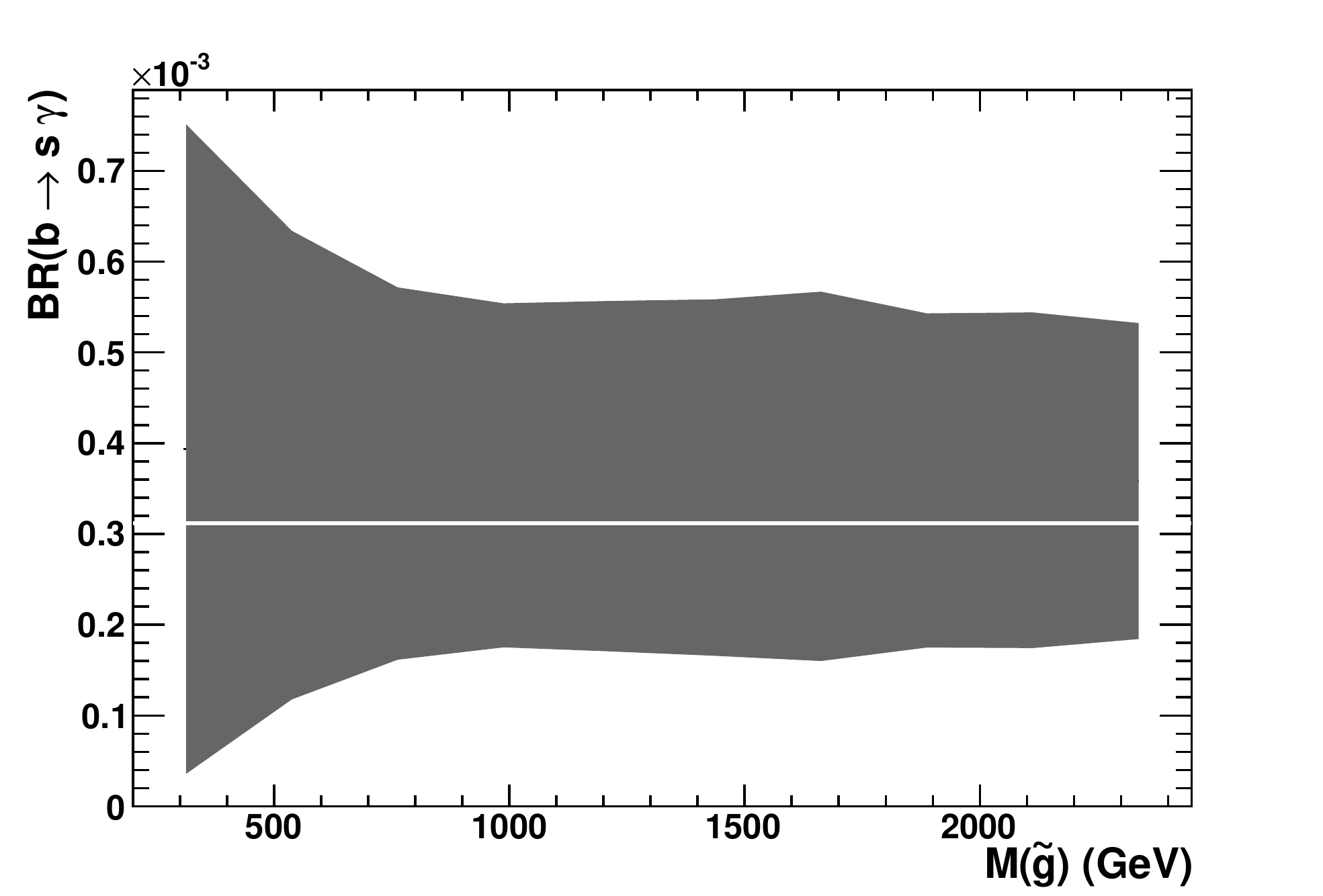} \\ 
\includegraphics[width=0.35\textwidth]{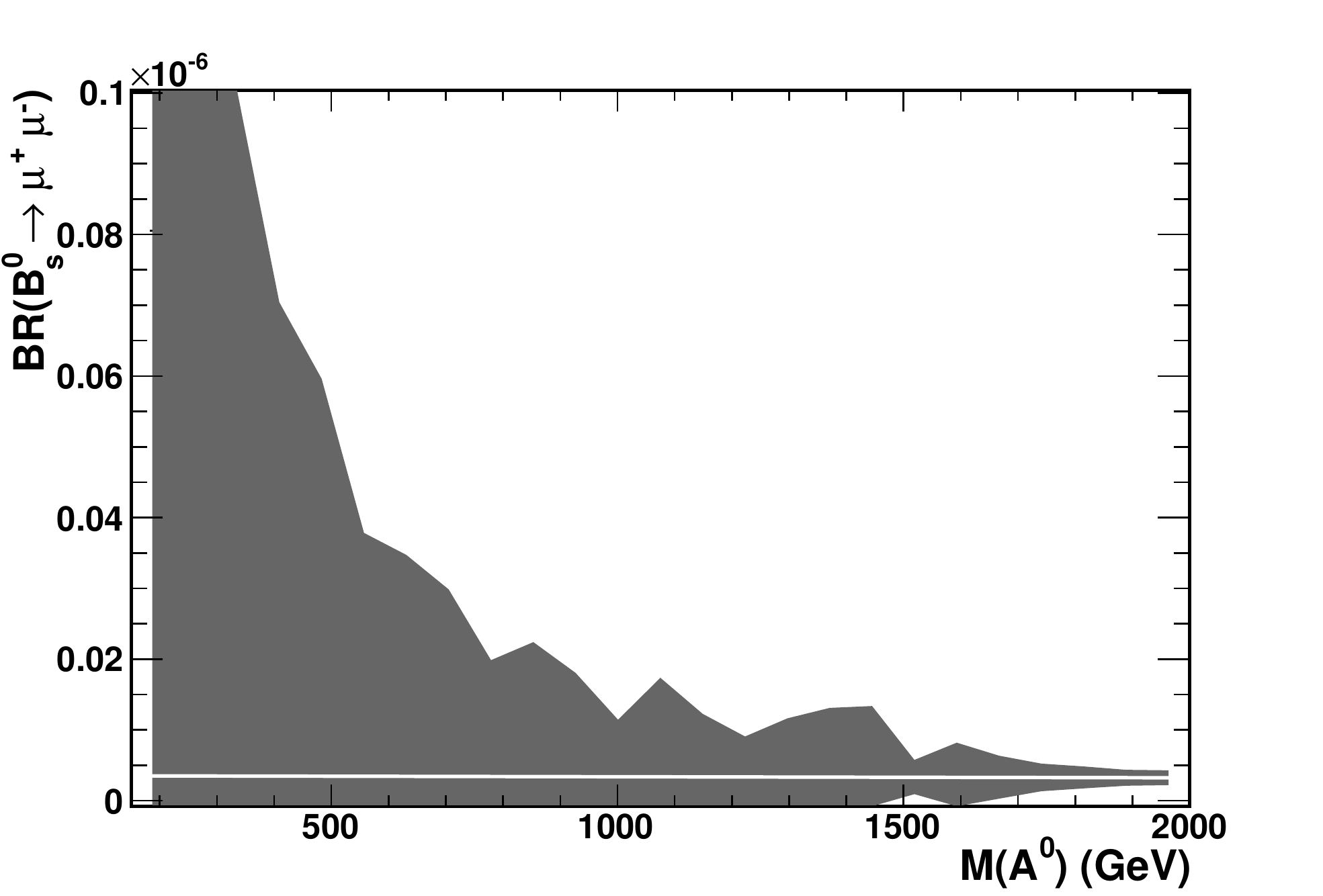}\\
\end{tabular}
\end{center}
\caption{Branching fractions of $B \to X_s \gamma$ (left panel) and $B_s \to \mu^+ \mu^-$ (right panel) of pMSSM points as a function of the gluino and $A^0$ boson masses, respectively. The white horizontal lines correspond to the SM values.}
\label{fig:bsg}
\end{figure}

The range of branching fractions of these processes in the pMSSM parameter space is displayed as a function of the gluino and the $A^0$ boson mass in Figure~\ref{fig:bsg}.

In addition to these rare processes, we consider a set of tree-level observables, which are very sensitive to the charged Higgs mass as well as to 
$\tan\beta$. These include $\mbox{BR}(B \to \tau \nu)$, $\mbox{BR}(D_s \to \tau \nu )$, $\mbox{BR}(B \to D \tau \nu)$ and $\mbox{R}_{\mu23}$ which is related 
to the $K \to \mu \nu$ decay. We use the following 95\% C.L. intervals, which include theoretical and experimental 
uncertainties~\cite{Mahmoudi:2008tp,Akeroyd:2009tn,Akeroyd:2010qy}:
\begin{eqnarray}
0.56 < &\displaystyle\frac{\mbox{BR}(B \to \tau \nu)}{\mbox{BR}_{SM}(B \to \tau \nu)}& < 2.70 \;,\\
4.7 \times 10^{-2} < &\mbox{BR}(D_s \to \tau \nu )& < 6.1 \times 10^{-2} \;,\\
2.9 \times 10^{-3} < &\mbox{BR}(B \to D^0 \tau \nu)& < 14.2 \times 10^{-3} \;,\\
0.985 < &\mbox{R}_{\mu23}(K \to \mu \nu) & < 1.013 \;.
\label{eq:leptonic}
\end{eqnarray}

We should emphasise here that the allowed intervals for the flavour observables are quite robust within the MSSM and we have taken into account consistently the theoretical and experimental uncertainties.

The anomalous magnetic moment of the muon, $g_\mu-2$, represents an outstanding open issue in precision low-energy measurements sensitive to SUSY 
contributions. The discrepancy between its SM prediction and the experimental value obtained by E821~\cite{Davier:2010nc,Hagiwara:2011af}, can be 
interpreted as a result of SUSY contributions~\cite{Martin:2001st}, which would favour light masses of scalar muons ruling out large region of 
the SUSY parameter space~\cite{Ellis:2007fu,Ellis:2001yu}. However, recent re-analyses of the hadronic contributions raise questions of the interpretation 
of the apparent discrepancy~\cite{Goecke:2010if,Boughezal:2011vw,Bodenstein:2011qy,Goecke:2011pe} and offer a mean to reconcile the experimental result 
with SM prediction. Given the remaining uncertainty around the interpretation of the $g_\mu-2$ result, we adopt here the broadest possible interval, which takes separately the lowest and highest 95\% C.L. bounds given in the recent literature: 
\begin{equation}
-2.4 \times 10^{-9} < \delta a_\mu < 4.5 \times 10^{-9} \;.
\end{equation}

The next observable we use as a constraint is the dark matter relic density $\Omega_{\mathrm{DM}} h^2$, for which we adopt the latest WMAP 
result~\cite{Komatsu:2010fb} from the cosmic microwave background analysis. It has been shown~\cite{Arbey:2008kv,Arbey:2009gt} that many cosmological 
phenomena can strongly alter the value of the calculated relic density, leading to a very broad interval, $10^{-4} < \Omega_{\mathrm{DM}} h^2 < 10^5$, 
which takes into account uncertainties from realistic cosmological scenarios~\cite{Arbey:2011gu}. Nevertheless, for this study we adopt the standard 
model of cosmology, and we consider that the lightest neutralino can account for the observed dark matter density from at least 0.1\% of the WMAP value, 
up to the WMAP central value increased by 3.5 times the WMAP uncertainty summed in quadrature with 10\% additional contribution to include theoretical uncertainties, in particular from the QCD equations of state in the relic 
density calculation~\cite{Hindmarsh:2005ix}. We therefore retain the following interval in our numerical analysis:
\begin{equation}
10^{-4} < \Omega_\chi h^2 < 0.155\;.
\end{equation}
In order to test the sensitivity of our analysis to the addopted range of these constraints, we increased the range of Eqs.~(\ref{eq:bsg})-(\ref{eq:leptonic}) by a factor 1.5 and repeat the analysis. Results are discussed in Sections~\ref{sec:4-1} and \ref{sec:4-2}.
In addition, the dark matter direct detection scattering cross-sections also provide relevant information. The CDMS \cite{Ahmed:2009zw} and the Xenon~100 \cite{Aprile:2011hi} collaborations have recently published stringent limits on the spin-independent elastic WIMP-nucleon scattering cross section. In this study, we calculate the scattering cross sections for each SUSY model point, but we do not apply it as a constraint. We discuss the implications of dark matter direct detection on our results in Section \ref{sec:4-4}.
\begin{figure}[t!]
\begin{center}
\includegraphics[width=0.42\textwidth]{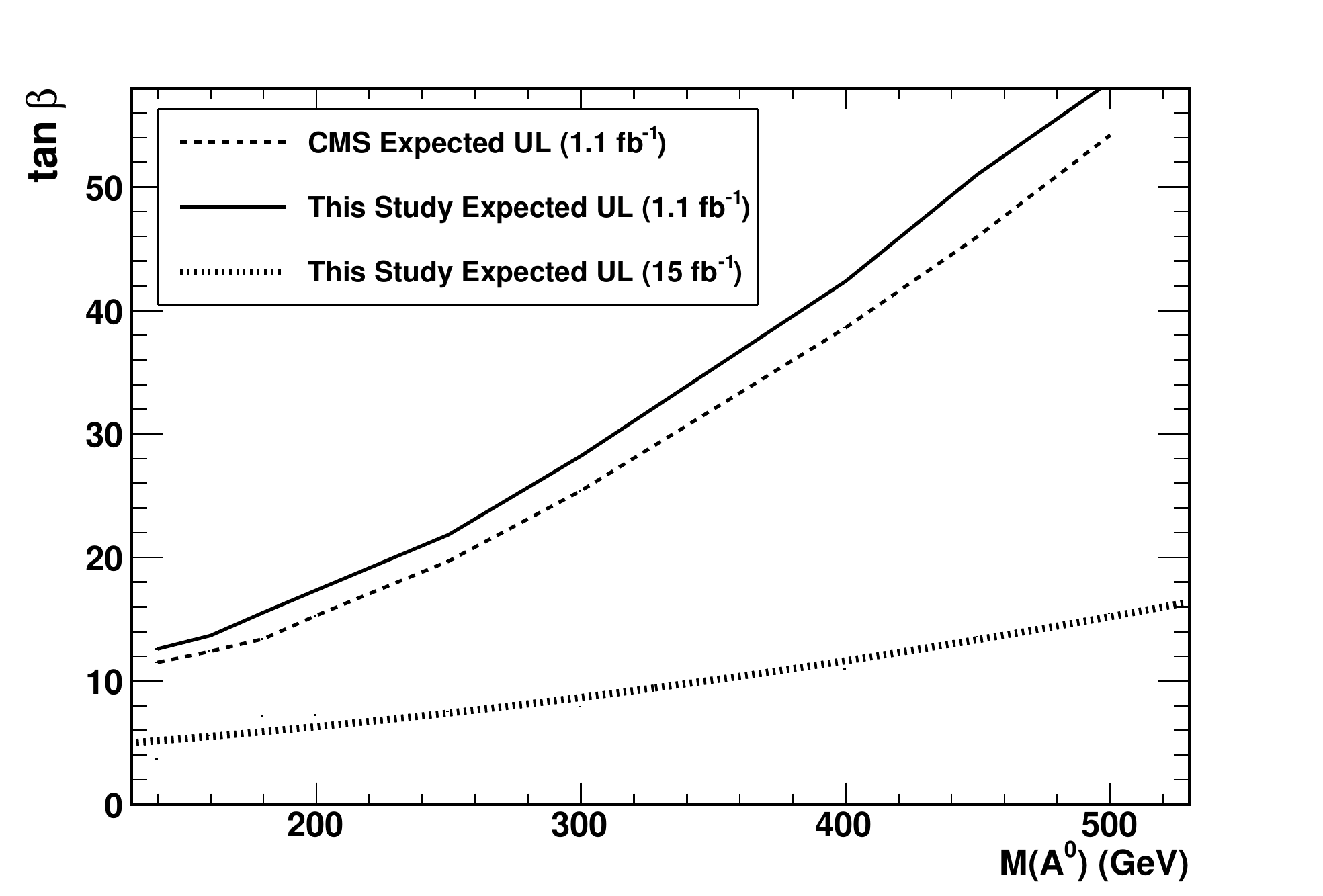}
\end{center}
\caption{95\% C.L. upper limit exclusion contours in the ($m_A,\tan\beta$) plane. The dashed line shows the expected exclusion from CMS with 1.1~fb$^{-1}$ \cite{11-009}, the solid line the corresponding contour obtained with our simulations and the dotted line the same contour with 15~fb$^{-1}$.}
\label{fig:mAtanb}
\end{figure}

Furthermore, we consider the constraints from direct Higgs searches at LEP, Tevatron and LHC, using the code {\tt HiggsBounds}, which incorporates also the recent results on light and heavy Higgs searches at LHC. In order to study the scaling of heavy Higgs exclusion in ($M_A,\tan\beta$) plane we estimate the sensitivity of the CMS analysis in the $\tau \tau$ channel \cite{11-009}. We calculate the product $\sigma_\phi \mathrm{BR}_{\tau\tau}$ as a function of $M_A$ and $\tan\beta$ in the NUHM model where we also vary $m_0$, $m_{1/2}$, $A_0$ and $\mu$. We reconstruct the wedge for the 95\% upper bound for 1.1~fb$^{-1}$ on $\tan\beta$ as a function of $M_A$ based on the expected sensitivity of the CMS analysis as given in \cite{11-009}. We then rescale the 95\% upper bound assuming an integrated luminosity of 15~fb$^{-1}$. The results are shown in Figure~\ref{fig:mAtanb}. We use this constraint for our study with 15~fb$^{-1}$ of data in Sections~\ref{sec:4-3} and \ref{sec:4-4}. 
Finally we impose the 
constraints on the SUSY masses summarised in Table~\ref{tab:constraintsSUSY}. In the present study we consider for simplicity a lower limit of 46~GeV for the lightest neutralino mass, even if this limit can be circumvented in specific cases.
\begin{table}
\begin{center}
\begin{tabular}{|c|c|c|}
\hline
Particle & Limits & ~~~~~~Conditions~~~~~~\\
\hline\hline
$\tilde \chi^0_1$ & 46 &  \\
\hline
$\tilde \chi^0_2$ & 62.4 & $\tan\beta < 40$\\
\hline
$\tilde \chi^0_3$ & 99.9 & $\tan\beta < 40$\\
\hline
$\tilde \chi^0_4$ & 116 & $\tan\beta < 40$\\
\hline
$\tilde \chi^\pm_1$ & 94 & $\tan\beta < 40$, $m_{\tilde \chi^\pm_1} - m_{\tilde \chi^0_1} > 5$ GeV\\
\hline
$\tilde{e}_R$ & 73 & \\
\hline
$\tilde{e}_L$ & 107 & \\
\hline
$\tilde{\tau}_1$ & 81.9 & $m_{\tilde{\tau}_1} - m_{\tilde \chi^0_1} > 15$ GeV\\
\hline
$\tilde{u}_R$ & 100 & $m_{\tilde{u}_R} - m_{\tilde \chi^0_1} > 10$ GeV\\
\hline
$\tilde{u}_L$ & 100 & $m_{\tilde{u}_L} - m_{\tilde \chi^0_1} > 10$ GeV\\
\hline
$\tilde{t}_1$ & 95.7 & $m_{\tilde{t}_1} - m_{\tilde \chi^0_1} > 10$ GeV\\
\hline
$\tilde{d}_R$ & 100 & $m_{\tilde{d}_R} - m_{\tilde \chi^0_1} > 10$ GeV\\
\hline
$\tilde{d}_L$ & 100 & $m_{\tilde{d}_L} - m_{\tilde \chi^0_1} > 10$ GeV\\
\hline
& 248 & $m_{\tilde \chi^0_1} < 70$ GeV, $m_{\tilde{b}_1} - m_{\tilde \chi^0_1} > 30$ GeV\\
& 220 & $m_{\tilde \chi^0_1} < 80$ GeV, $m_{\tilde{b}_1} - m_{\tilde \chi^0_1} > 30$ GeV\\
$\tilde{b}_1$ & 210 & $m_{\tilde \chi^0_1} < 100$ GeV, $m_{\tilde{b}_1} - m_{\tilde \chi^0_1} > 30$ GeV\\
& 200 & $m_{\tilde \chi^0_1} < 105$ GeV, $m_{\tilde{b}_1} - m_{\tilde \chi^0_1} > 30$ GeV\\
& 100 & $m_{\tilde{b}_1} - m_{\tilde \chi^0_1} > 5$ GeV\\
\hline
$\tilde{g}$ & 195 & \\
\hline
\end{tabular}
\end{center}
\caption{Constraints on the SUSY particle masses (in GeV) from searches at LEP and the Tevatron~\cite{Nakamura:2010zzi}.
\label{tab:constraintsSUSY}}
\end{table}

\begin{figure*}[ht!]
\begin{center}
\begin{tabular}{c}
\includegraphics[width=0.42\textwidth]{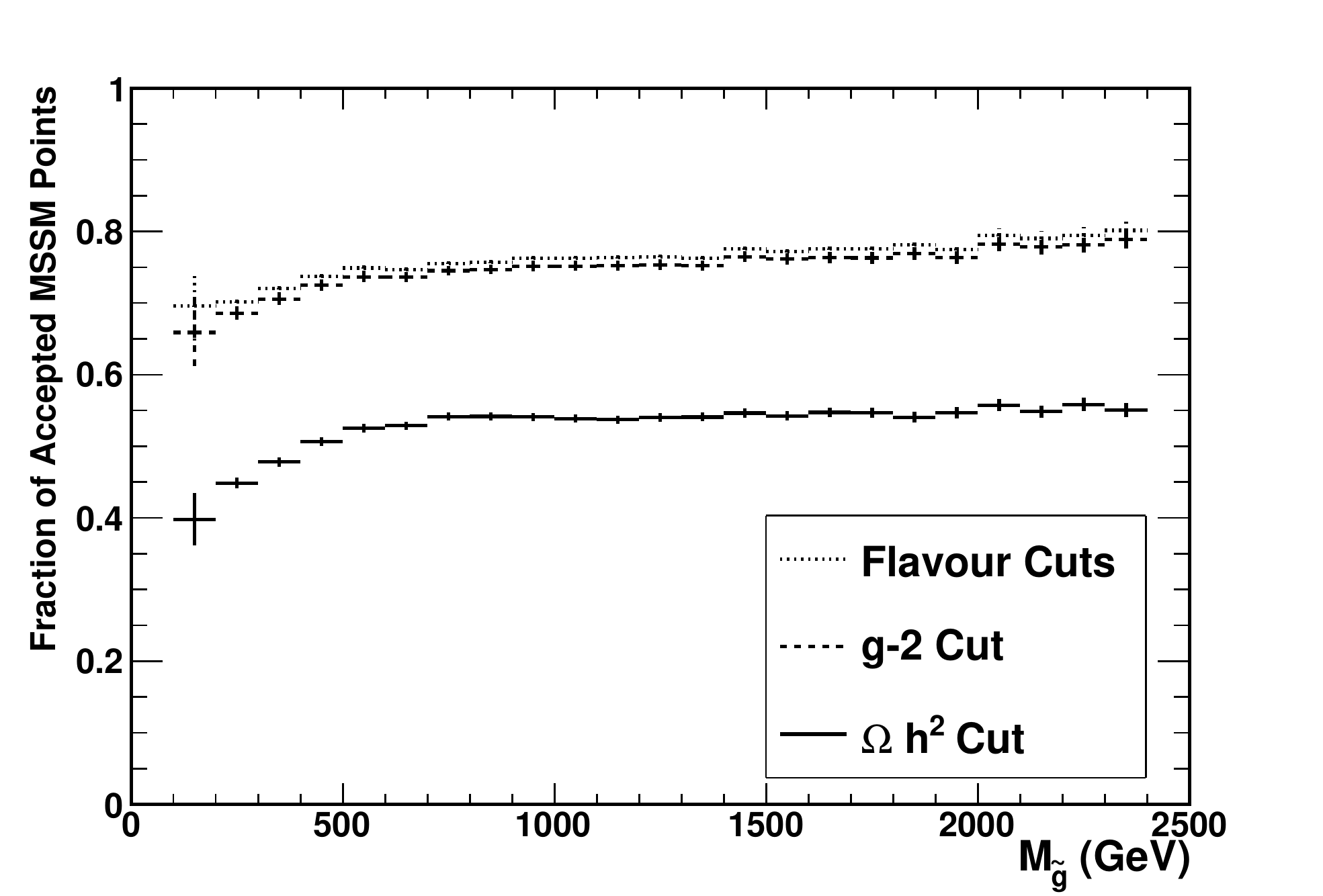}
\includegraphics[width=0.42\textwidth]{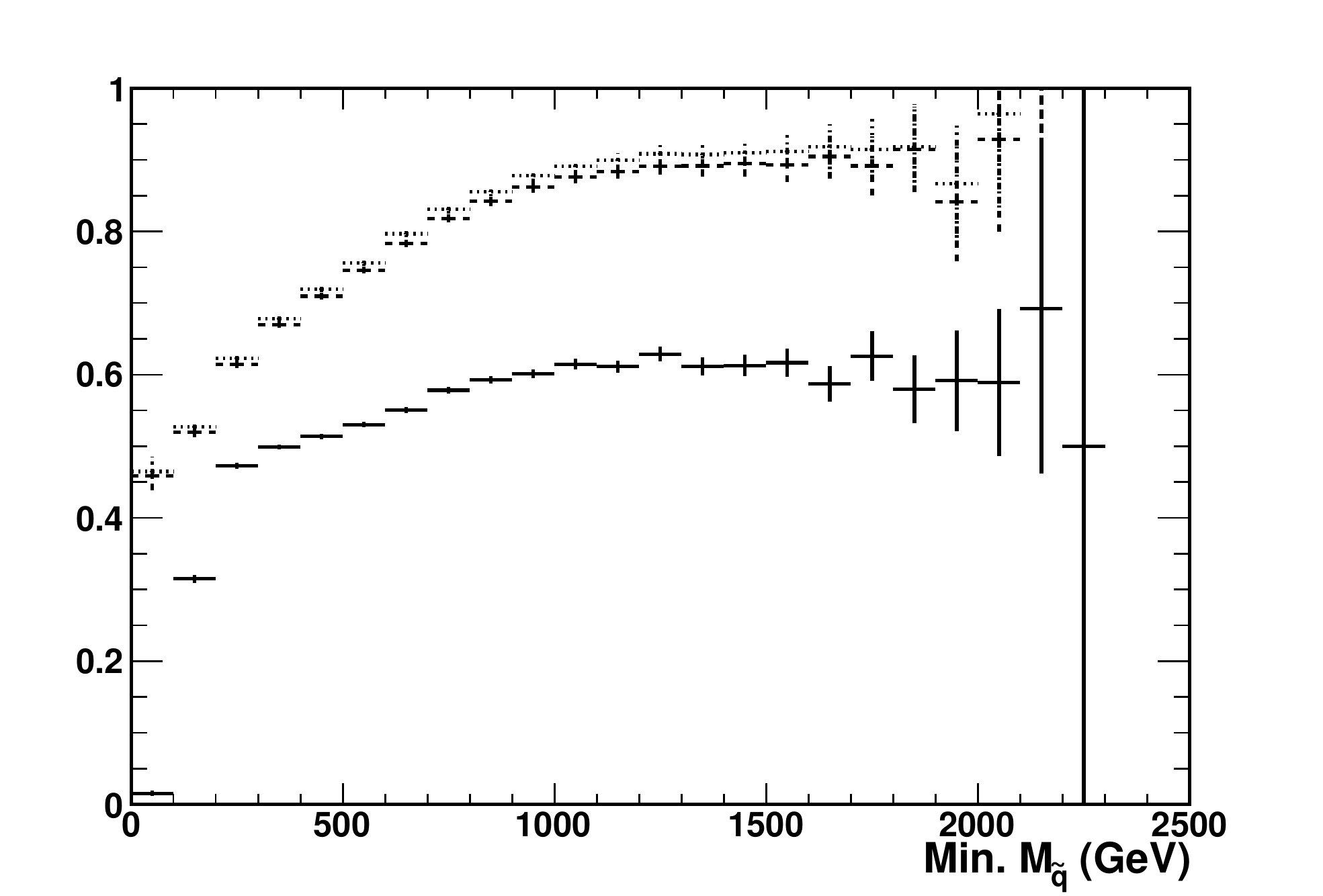} \\
\end{tabular}
\end{center}
\caption{Fraction of valid MSSM points selected after the flavour (dotted line), $g_\mu-2$ (dashed line) and $\Omega_\chi h^2$ 
(solid line) subsequent constraints as a function of the masses of strongly-interacting SUSY particles: the gluino 
(upper panel) and the lightest scalar quark (lower panel).}
\label{fig:sconst}
\end{figure*}

\begin{figure*}[ht!]
\begin{center}
\begin{tabular}{c}
\includegraphics[width=0.42\textwidth]{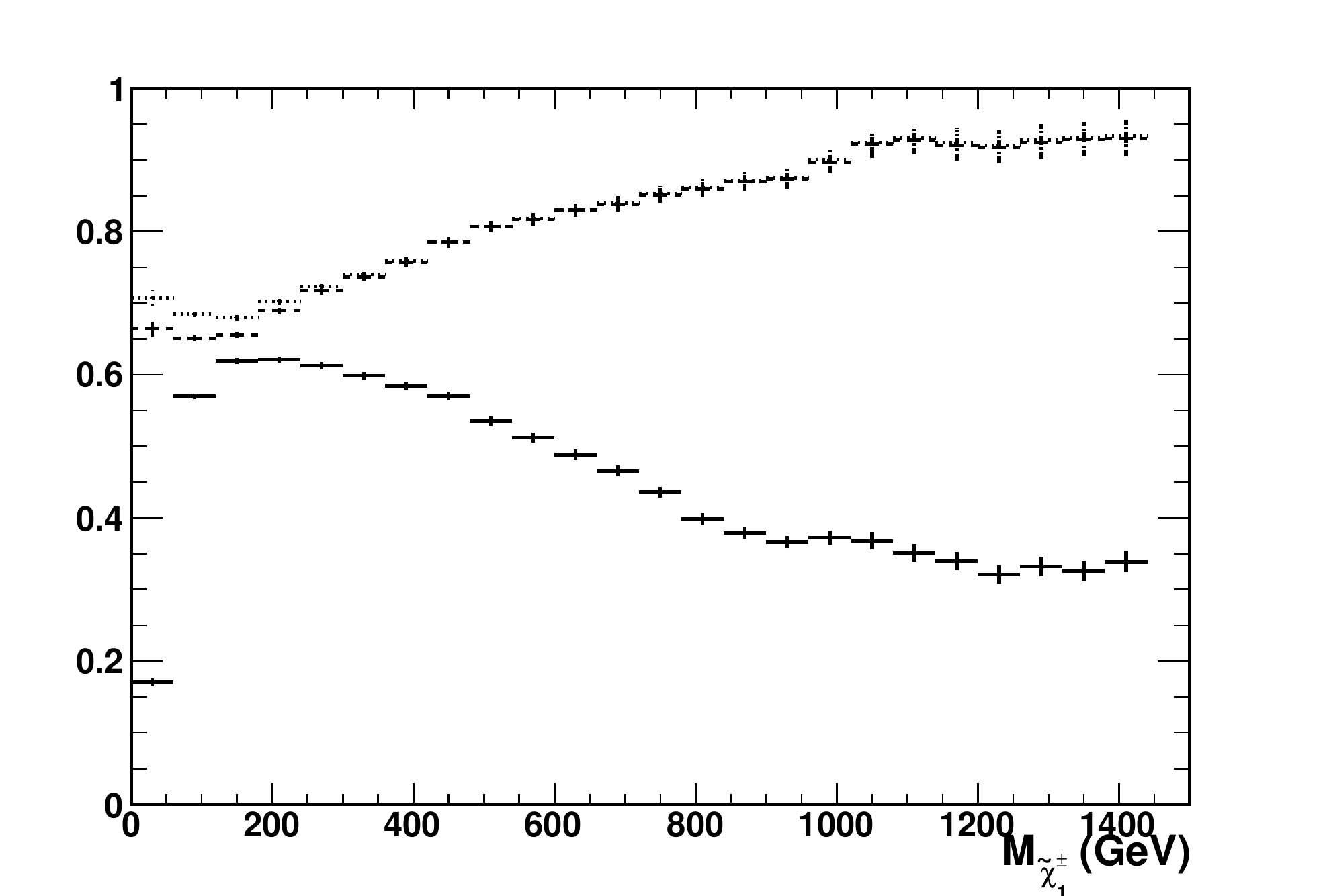}
\includegraphics[width=0.42\textwidth]{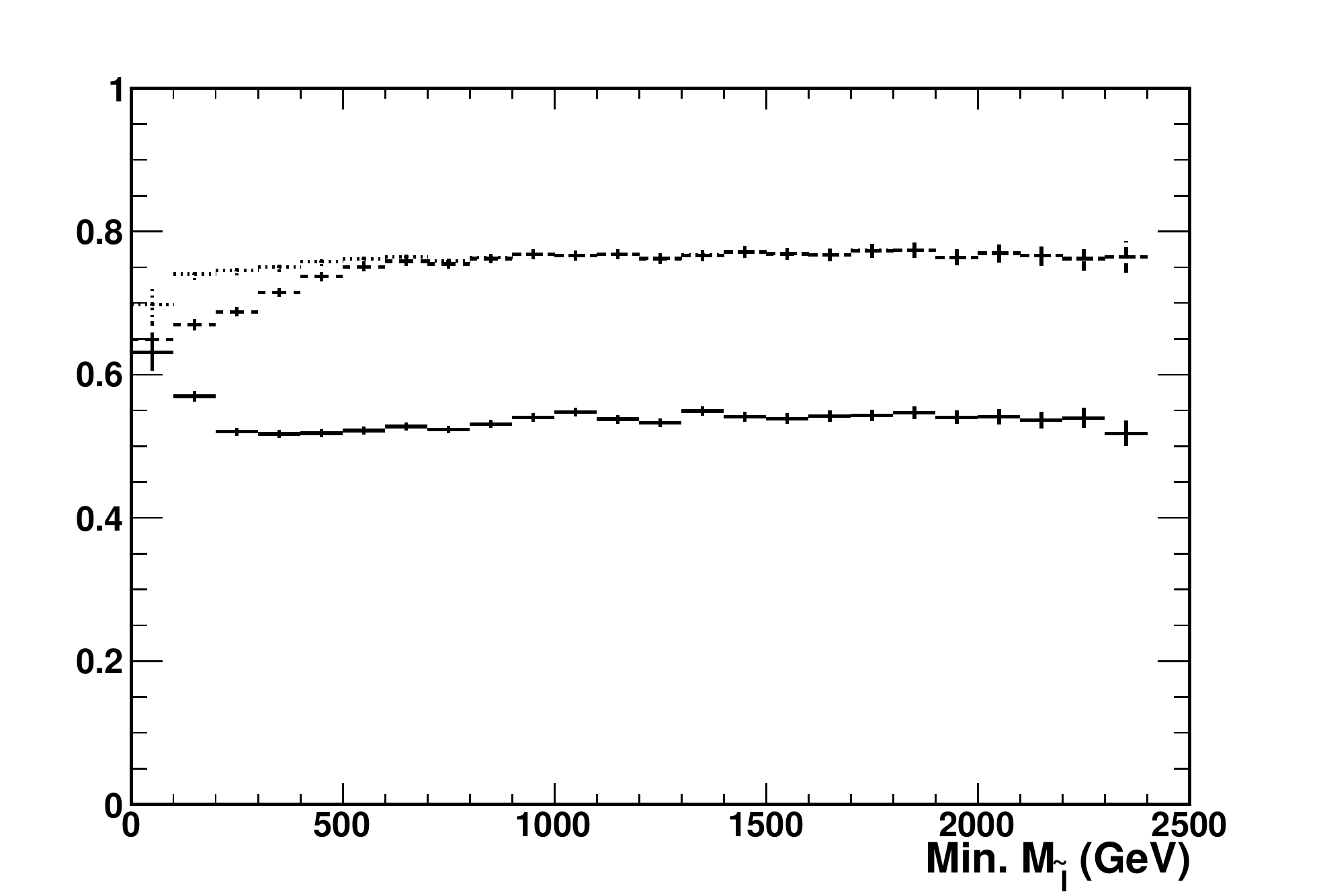} \\
\end{tabular}
\end{center}
\caption{Fraction of valid MSSM points selected after the flavour (dotted line), $g_\mu-2$ (dashed line) and $\Omega_\chi h^2$ 
(solid line) subsequent constraints as a function of the masses of two weakly-interacting SUSY particles: the 
lightest chargino $\tilde \chi^{\pm}_1$ (upper panel) and the scalar lepton (lower panel).}
\label{fig:wconst}
\end{figure*}

The effect of these constraints on the masses of some SUSY particles is shown in Figures~\ref{fig:sconst} and 
\ref{fig:wconst}, where the fraction of valid MSSM points fulfilling the subsequent cuts on flavour, $g_\mu-2$ 
and $\Omega_\chi h^2$ observables is shown as a function of the masses of $\tilde g$, lightest $\tilde q$, 
$\tilde \chi^{\pm}_1$ and $\tilde \ell$ ($\ell$ = $e$, $\mu$).
We observe that the flavour cuts disfavour masses below $\sim$500~GeV for the gluino and the lightest scalar quark due 
to the deviations arising from their contributions to heavy flavour loop-mediated decays. Very light, as well as heavy charginos are equally disfavoured by the relic dark matter constraint, which slightly favours light sleptons. However, 
with the exception of gluinos and lightest $\tilde q$ below 250~GeV and lightest charginos of more than 1~TeV, the 
acceptance of all these cuts appear quite flat with the mass of both strongly and weakly interacting SUSY particles.

\section{LHC analysis simulation}
\label{sec:3}
The sensitivity of the LHC SUSY searches with jets, leptons and missing energy for the pMSSM points selected from our 
scans has been assessed by reproducing on fast simulation the event selection criteria of three analyses of the CMS 
Collaboration. These analyses are the SUSY searches in all hadronic events with $\alpha_T$~\cite{11-003}, in same-sign 
isolated dilepton events with jets and missing energy~\cite{11-010} and in opposite-sign dilepton events with missing 
transverse energy~\cite{11-011}. Since these analyses are cut based, the observation or exclusion of a given model 
is obtained from the number of events observed in the signal region and the events selected by each analysis and signal 
regions are fully uncorrelated, they are well suited for a fast simulation and reconstruction study.
While we have adopted the CMS analyses for this study, the ATLAS experiment has comparable sensitivity to SUSY.

Events generated with {\tt PYTHIA} are subsequently reconstructed and classified following the procedure of the three 
CMS analyses from the physics objects obtained from the {\tt Delphes} fast simulation. Jets are reconstructed using the 
anti-kt algorithm~\cite{Cacciari:2008gp}, implemented in the {\tt FastJet} package~\cite{Cacciari:2005hq}. The hadronic $\alpha_T$ analysis 
considers events fulfilling the selection requirements of ref.~\cite{11-003} binned according to their $H_T$ value. 
The four bins:
\begin{itemize}
 \item 575 $< H_T <$ 675~GeV,
 \item 675 $< H_T <$ 775~GeV,
 \item 775 $< H_T <$ 875~GeV,
 \item $H_T >$ 875~GeV,
\end{itemize}
are used to 
search for the SUSY signal. The same-sign dilepton analysis considers high-$p_t$ leptons in the four search 
regions defined by
\begin{itemize}
 \item $H_T >$ 400~GeV and $MET >$ 120~GeV, 
 \item $H_T >$ 400~GeV and $MET >$ 50~GeV, 
 \item $H_T >$ 200~GeV and $MET >$ 120~GeV, 
 \item  $H_T >$ 80~GeV and $MET >$ 100~GeV. 
\end{itemize}
The opposite-sign dilepton  analysis looks to the high $MET$ and high $H_T$ search regions defined by
\begin{itemize}
 \item  $MET >$ 275~GeV and $H_T >$ 300~GeV, 
 \item $MET >$ 200~GeV and $H_T >$ 600~GeV, 
\end{itemize}
respectively. 
The expected number of background events from SM processes in each search region is taken from the published CMS results, 
obtained with full simulation and reconstruction, validated on data, and rescaled to the assumed integrated luminosity. 
The 95\% confidence level exclusion of each SUSY point in presence of background only is determined using the CLs 
method~\cite{Read:2002hq}. 
The number of events in each of the search regions of the CMS analyses described above is computed for the signal plus 
background and the background only hypotheses, where the signal represents the number of events in the search regions 
estimated for the SUSY point under test. The number of events in each of the search regions is assumed to be uncorrelated. 

\subsection{Simulation Validation}
\label{sec:3-1}
{\color{black}In order to ensure that the sensitivity to signal events in our fast simulation and reconstruction is representative of 
that from the analysis of the real CMS data, a validation of this procedure is performed comparing results of the fast 
simulation used in this analysis to those of signal events fully simulated and reconstructed with the CMS code. 
These events are chosen from the set of LM benchmark points, defined in the CMSSM model and used by the experiment 
for studies of the analysis sensitivities \cite{Ball:2007zza}. The fast simulation and reconstruction validation is performed 
in three stages. 
First, we compare the shape of event variables after event preselection. These are the number of reconstructed jets, the $\alpha_T$ 
variable used to discriminate against QCD jets and the sum of the jet transverse energies $H_T$ for the hadronic analysis; 
the $p_T$ of the two leading leptons, the dilepton invariant mass, the event missing transverse 
energy, $MET$, and $H_T$ variable for the dilepton channels. The shape of the distributions in the CMS simulation and reconstruction 
and in the {\tt Delphes} fast simulation are in good agreement. Then, we compare the number of signal events selected in the different 
signal regions for each LM point. 
\begin{figure*}[t!]
\begin{center}
\begin{tabular}{c}
\includegraphics[width=0.26\textwidth]{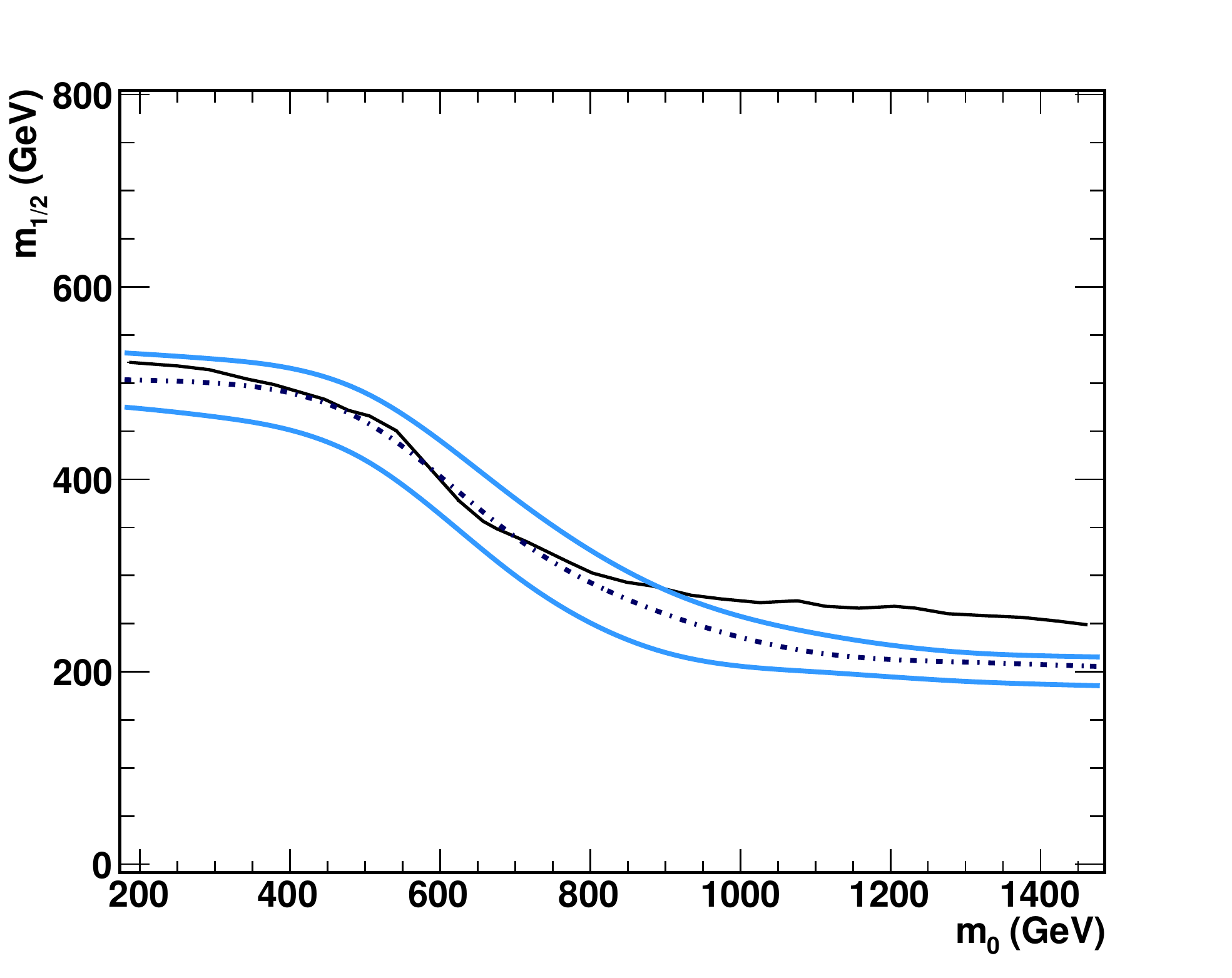}
\includegraphics[width=0.31\textwidth]{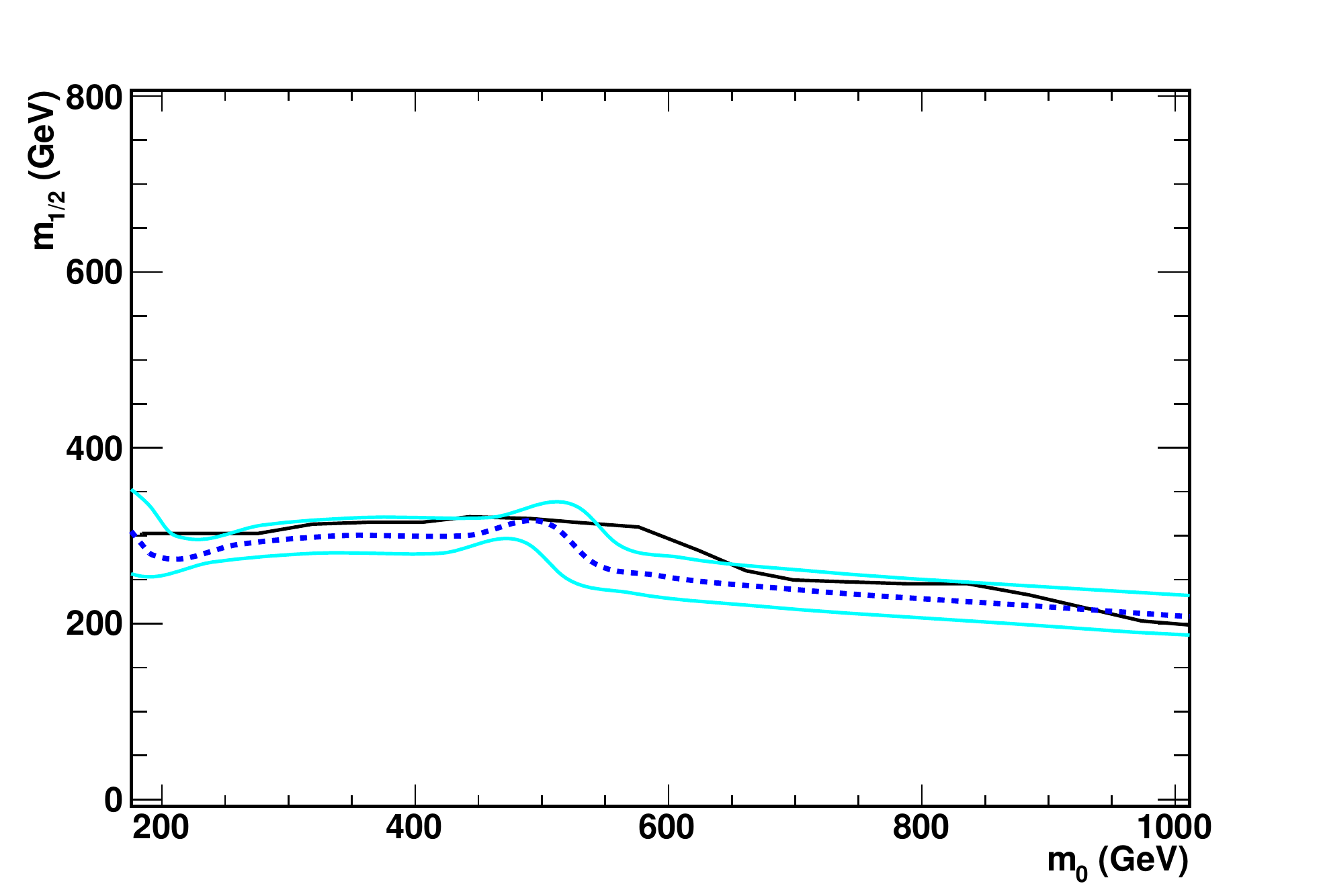}
\includegraphics[width=0.31\textwidth]{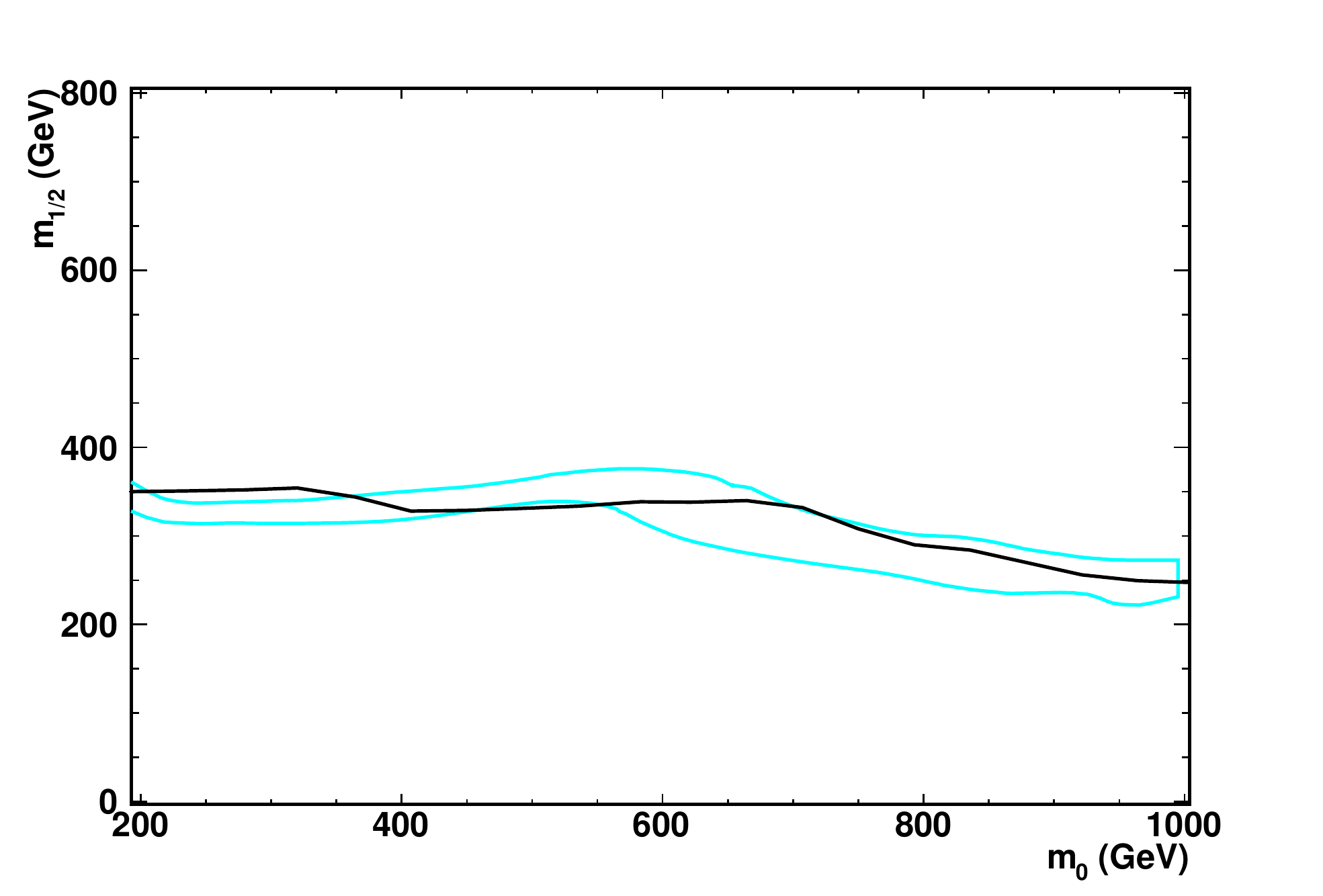} \\
\end{tabular}
\end{center}
\caption{95\% C.L. exclusion contours in the ($m_0$, $m_{1/2}$) plane for CMSSM points with $\tan \beta$ = 10, $A_0$ = 0 
and sign($\mu$) = +1 for 1~fb$^{-1}$ of data at 7~TeV. The contours obtained from our fast simulation analysis (black solid line) are 
compared to the bands of the expected exclusion limit from the CMS full simulation and reconstruction including uncertainties shown by 
the lighter coloured lines.}
\label{fig:cmssm-validation}
\end{figure*}
In general, the agreement between CMS full simulation and reconstruction and the {\tt Delphes} fast simulation is good, with differences in 
the observed signal event rates below 25\%. Finally, we compare
the exclusion contours obtained using our fast simulation in the ($m_0$, $m_{1/2}$) plane of the CMSSM model for the three 
analyses with the expected exclusion regions reported by CMS. Again, the 95\% C.L. exclusion contours for the $m_0$ and 
$m_{1/2}$ CMSSM parameters obtained with our simulation agree within 20\% to those of CMS as shown in 
Figure~\ref{fig:cmssm-validation}. 
}

\section{Results}
\label{sec:4}
Results are presented for 1~fb$^{-1}$ and 15~fb$^{-1}$ of integrated luminosity, corresponding 
to the statistics reported by CMS to the EPS 2011 conference and that expected by the end of 
2012, respectively. These are based on the analysis of the 835k pMSSM points fulfilling all the selection criteria discussed above and for which events have been successfully simulated, selected 
out of 24.57M pMSSM generated points.  The statistics and efficiency if the subsequent selection 
steps, from the generated points giving a valid {\tt SOFTSUSY} 
spectrum through the mass, low-energy data and $\Omega_\chi h^2$ constraints to those for which a 
sample of events for inclusive SUSY production at 7~TeV is successfully simulated with {\tt PYTHIA} 
are summarised in Table~\ref{tab:stat}. 
We discuss results for strongly interacting SUSY particles ($\tilde g$ and $\tilde q$), to which 
the present LHC searches are directly sensitive, 
and for their weakly interacting companions ($\tilde \ell$ and $\tilde \chi$), which are more 
indirectly probed at the LHC and are instead of major interest for the planning of a future 
high energy lepton collider. 
We present results in terms of the fraction of pMSSM points compatible with the 
constraints discussed in Section~\ref{sec:2-2} (referred to as ``accepted pMSSM points'' in the following) which are excluded by 
the LHC searches as a function of the mass of the SUSY particle of interest. 
\begin{table}
\begin{center}
\begin{tabular}{|l|c|c|c|}
\hline
Selection         & pMSSM     &  Selection  &  Cumulative \\
                  & points   &  Efficiency &  Efficiency \\
\hline
\hline
Generated Points    & 24.57M  & 1     & 1 \\
\hline
Valid Spectra       & ~9.41M  & 0.383 & 0.383 \\
\hline
$\tilde \chi^0_1$ LSP and      &        &       &  \\
Mass Limits         & ~2.62M  & 0.278 & 0.107 \\
\hline
Higgs Limits        & ~1.81M  & 0.691 & 0.074 \\
\hline
Flavour and $g_\mu-2$  & ~1.34M  & 0.743 & 0.055 \\
\hline
$\Omega_\chi h^2$ & ~897k  & 0.668 & 0.037 \\
\hline
Successfull         &        &       &  \\
Simulation          & ~835k  & 0.931  &  0.034 \\
\hline
\end{tabular}
\caption{Scan statistics. Points for which a 7~TeV $pp$ inclusive SUSY event sample is successfully
simulated and reconstructed are used in the subsequent analysis.}
\end{center}
\label{tab:stat}
\end{table}
We study the scan coverage in terms of the characteristics of the SUSY spectra and particle properties. We observe that in 33.2\% of accepted points, the gluino is lighter than the lightest of the squarks of the first two generations. 13.3\% of the points have the gluino heavier than any of the squarks. Compressed spectra are particularly challenging for searches at the LHC. We observe that only 1.5\% of our accepted points have $M_{\tilde g}-M_{\chi^0_1} <$ 50~GeV, while 3.7\% have $M_{\tilde q}-M_{\chi^0_1} <$ 50~GeV, where $\tilde q$ is lightest squark of the first two generations. Turning to the particle properties, the Higgs boson mass ranges from 80 to 135~GeV. The lightest neutralino is bino-like in 6\%, wino-like in 23\%, and higgsino-like in 24\% of the accepted points.

Further, we study the sensitivity of the three analyses used. We observe that of the points excluded with 1 (15) fb$^{-1}$ of data, 96.2 (96.9)\% are excluded by the fully hadronic analysis, 27.6 (30.0)\% by the same sign lepton and 10.4 (16.1)\% by the opposite sign lepton analysis. 3.8 (3.1)\% of the points are excluded by leptonic analyses but not by the hadronic analysis.

\subsection{Strongly Interacting Sparticle Spectra of Allowed pMSSM Points}
\label{sec:4-1}
\begin{figure}[t!]
\begin{center}
\begin{tabular}{c}
\includegraphics[width=0.42\textwidth]{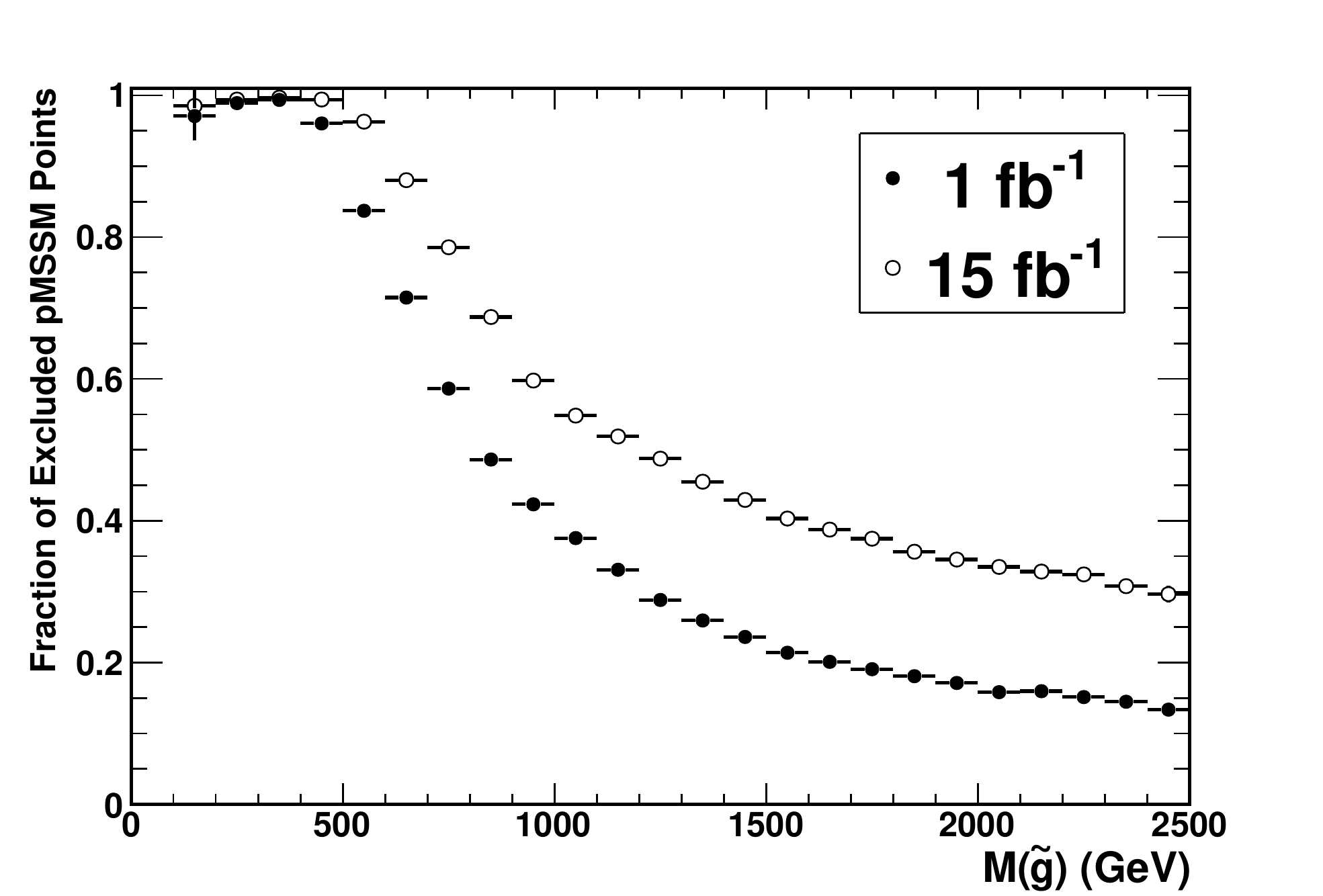} \\
\includegraphics[width=0.42\textwidth]{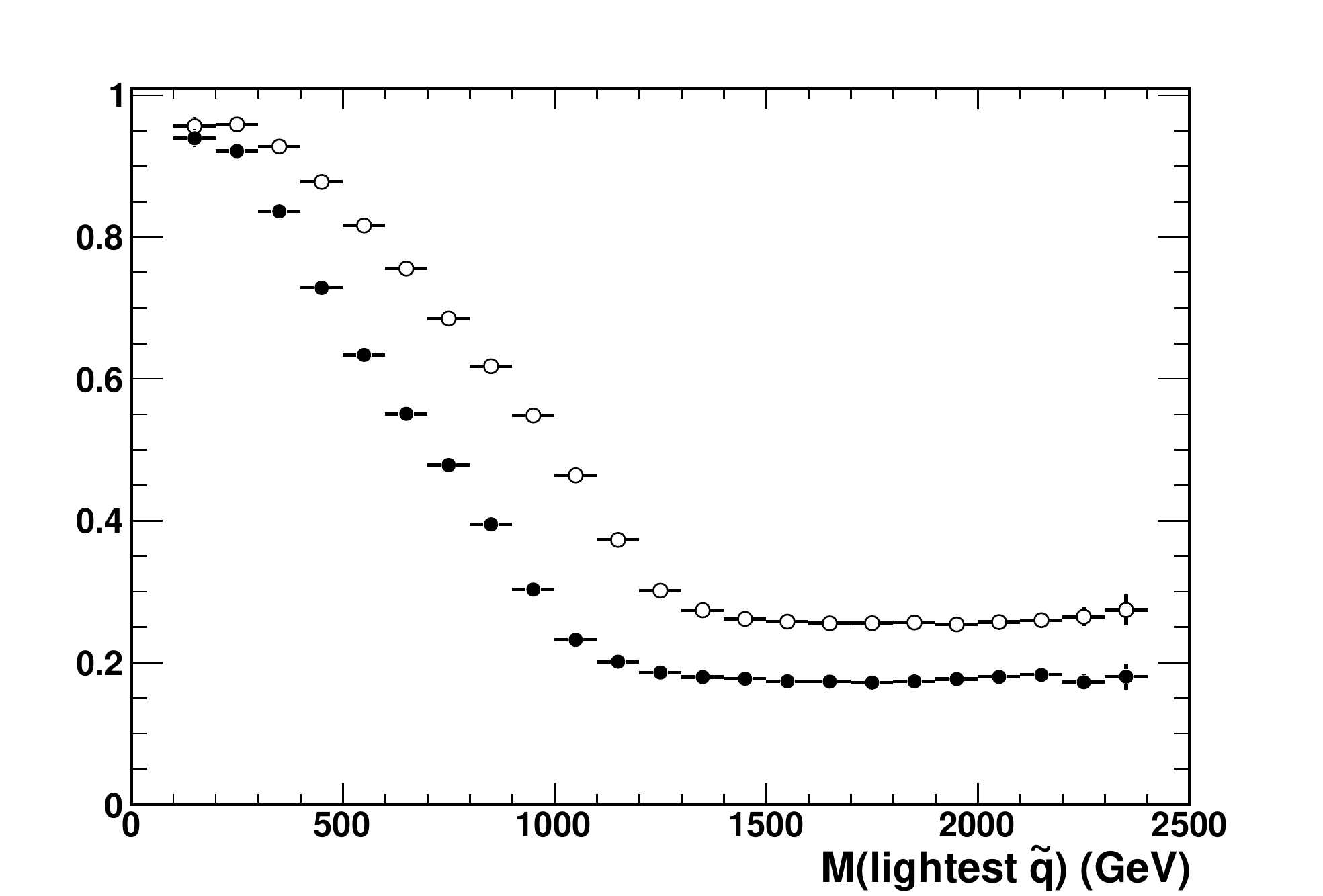} \\
\end{tabular}
\end{center}
\caption{Fraction of valid MSSM points excluded at 95\% C.L. by 1~fb$^{-1}$ and 15~fb$^{-1}$ of LHC data as a function of the masses of 
the gluino $\tilde g$ (upper panel) and the lightest scalar quark of the first two generations (lower panel).}
\label{fig:s1fb}
\end{figure}

\begin{figure}[t!]
\begin{center}
\includegraphics[width=0.42\textwidth]{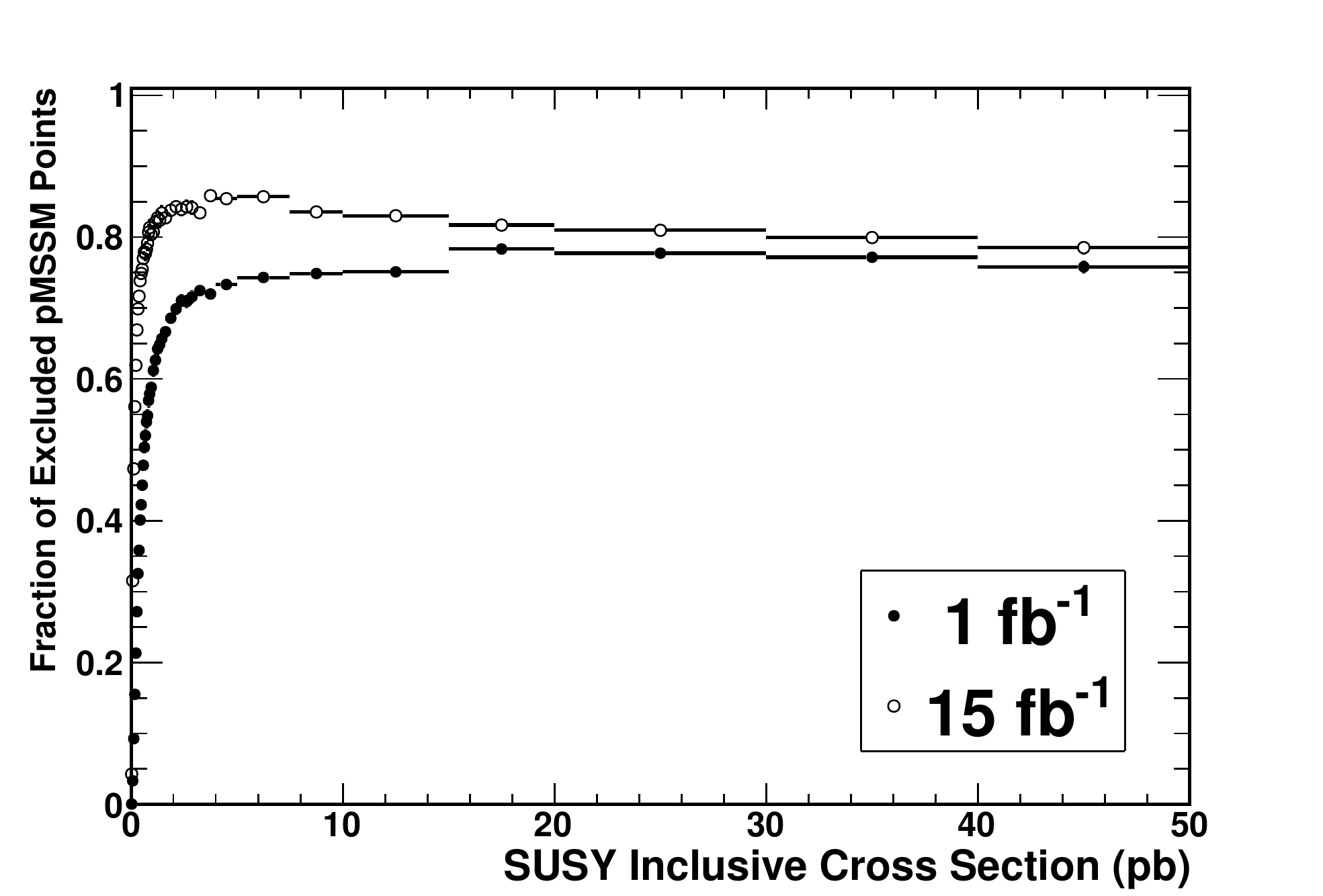}
\end{center}
\caption{Fraction of valid MSSM points excluded at 95\% C.L. by 1~fb$^{-1}$ and 15~fb$^{-1}$ of LHC data as a function of the inclusive 
SUSY cross section.}
\label{fig:xsec}
\end{figure}

We consider the masses of the gluino, $\tilde g$ and of the lightest scalar quark of the first two generations.
Figure~\ref{fig:s1fb} shows the fractions of accepted pMSSM points, which 
can be excluded by the results of the CMS analyses on 1~fb$^{-1}$ and 15~fb$^{-1}$ of data. We observe that in 
the case of the gluino the LHC data can exclude more than 85\% of the pMSSM points up to a mass of $\sim$520~GeV 
and 700~GeV for 1~fb$^{-1}$ and 15~fb$^{-1}$, respectively. 
The small fraction of points escaping exclusion below these mass limits have either small production cross section or small mass difference between the gluino and the LSP, resulting in compressed SUSY spectra yielding a lower transverse energy for the final state, as already pointed out in \cite{Conley:2011nn}. Above these masses, the sensitivity of the 
LHC data decreases rather sharply and disappears for $M_{\tilde g} \gtrsim$ 1.5~TeV. For the case 
of the lightest scalar quark of the first two generations, the points which can be excluded by the LHC data account for more than 85\% of the 
accepted pMSSM points up to mass values of 320~GeV and 510~GeV for 1~fb$^{-1}$ and 15~fb$^{-1}$ of data, respectively, 
while the effect of the LHC data disappears above 1.0~TeV and 1.3~TeV.
These curves provide us with generic mass limits for strongly interacting SUSY particles which extend the results obtained 
on specific models, such as the CMSSM. As a comparison, in the CMSSM the limit on the gluino mass for 1~fb$^{-1}$ of 7~TeV 
data is between 500~GeV and 1.1~TeV depending on $m_0$. In general, there are set of pMSSM parameters corresponding to 
spectra which are not detectable with the CMS analyses discussed above. Figure~\ref{fig:xsec} shows the fraction of 
valid MSSM points excluded at 95\% C.L. by 1~fb$^{-1}$ and 15~fb$^{-1}$ of LHC data as a function of the inclusive SUSY cross section for 
7~TeV $pp$ collisions. Even at values of the SUSY cross section in excess of 5~pb which accounts for 19\% of accepted points, approximately 20\% of the pMSSM points 
are not observables because of the low transverse energy in the SUSY decay products. These points deserve a special attention in future LHC studies.
pMSSM points with very compressed spectra having a mass difference between the gluino or the lightest of the scalar quarks of the first two generations and the LSP below the jet $p_T$ cut used in the analyses account for 5\% of the total accepted points and 6\% of those not excluded but having a SUSY inclusive cross section larger than 5~pb. 

We assess the sensitivity of our results on the adopted range for the input SUSY parameters and the constraints by studying the change of the fractions of accepted pMSSM points excluded with 1~fb$^{-1}$ of data and $M_{\tilde g}< 500$ and $M_{\tilde q}<400$ GeV on a dedicated scan of 1M generated points. These fractions are 97.8\% and 89.9\% for our standard ranges and become 95.3\% and 83.4\% if we increase the SUSY parameter range by a factor 1.5. This shows that the results of these analyses in the range of sensitivity of the LHC data are only moderately sensitive to the chosen range of SUSY parameters.
If we make looser cuts for constraints (\ref{eq:bsg})-(\ref{eq:leptonic}) corresponding to 3.5 $\sigma$ range the fraction of points passing this cut increases from 76\% to 85\%, and the fractions above become 95.6\% and 83.3\% for gluinos and squarks respectively. Finally, we apply tighter relic dark matter density cut corresponding to the 95\% C.L. of the seven year WMAP result ($0.0924 < \Omega_\chi h^2 < 0.1316$). This reduces the fraction of accepted points from 66.8\% to 0.7\% and the fractions of excluded points in the selected gluino and squark mass range become 99.8\% and 95.0\%.

\subsection{Weakly Interacting Sparticle Spectra of Allowed pMSSM Points}
\label{sec:4-2}

Likelihood analyses of constrained SUSY models have indicated that, as 
a result of the LHC searches, which have excluded the region of 
parameters favoured by data before the start of the LHC, the masses of 
SUSY particles over the allowed portion of the parameter space have 
shifted to larger values \cite{Buchmueller:2011sw}. 
Instead, the impact of the LHC searches for gluinos and scalar quarks on 
weakly interacting supersymmetric particles in the general MSSM is only 
indirect. In the absence of the mass relations between particles of the 
strongly- and weakly-interacting sectors, typical of constrained models 
such as the CMSSM or the NUHM, the masses of gauginos and sleptons in the 
MSSM are {\sl a priori} uncorrelated to those of the gluino and the squarks,
since their mass parameters are free and independent. 

Correlations are introduced either by the constraints applied from flavour 
physics and relic dark matter density or by the signatures in the LHC analyses. 
A correlation between the $\tilde \chi^{\pm}_1$ and the $\tilde q$ masses 
is observed in this study and originates through the dark matter relic density constraint. 
Correlations between the masses of the $\tilde q$ and those of 
$\tilde \chi^{\pm}$, $\tilde \chi^{0}$,  $\tilde \ell^{\pm}$ 
arise from searches in final states with leptons and missing energy. 
These topologies arise from cascade decays involving gauginos and sleptons, 
such as $\tilde q \to q \tilde \chi$, $\tilde \chi \to \ell \tilde \ell$, 
$\tilde \ell \to \ell \tilde \chi^0_1$. The negative result in searches with 
these topologies suppresses pMSSM points with $\tilde \chi$ and $\tilde \ell$ 
with large enough mass splittings from the $\tilde q$ to give large transverse energies to the leptons searched in the analyses.  
As a result of these correlations, the negative result in the search for 
strongly interacting supersymmetric particles may allow us to make an 
inference on the masses of gauginos and sleptons in the allowed region of 
the parameter space. 
\begin{figure*}[ht!]
\begin{center}
\begin{tabular}{c}
\includegraphics[width=0.42\textwidth]{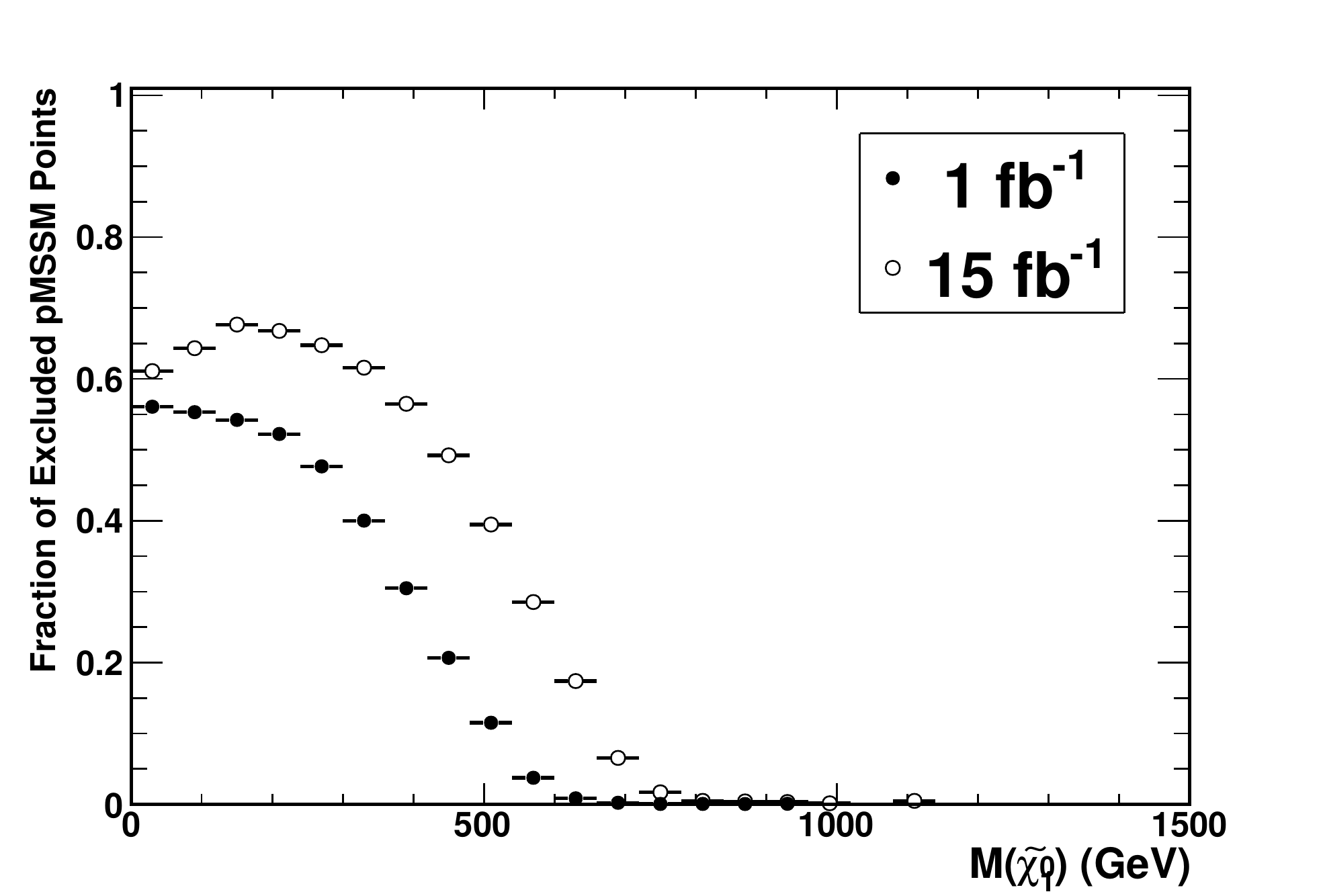}\includegraphics[width=0.42\textwidth]{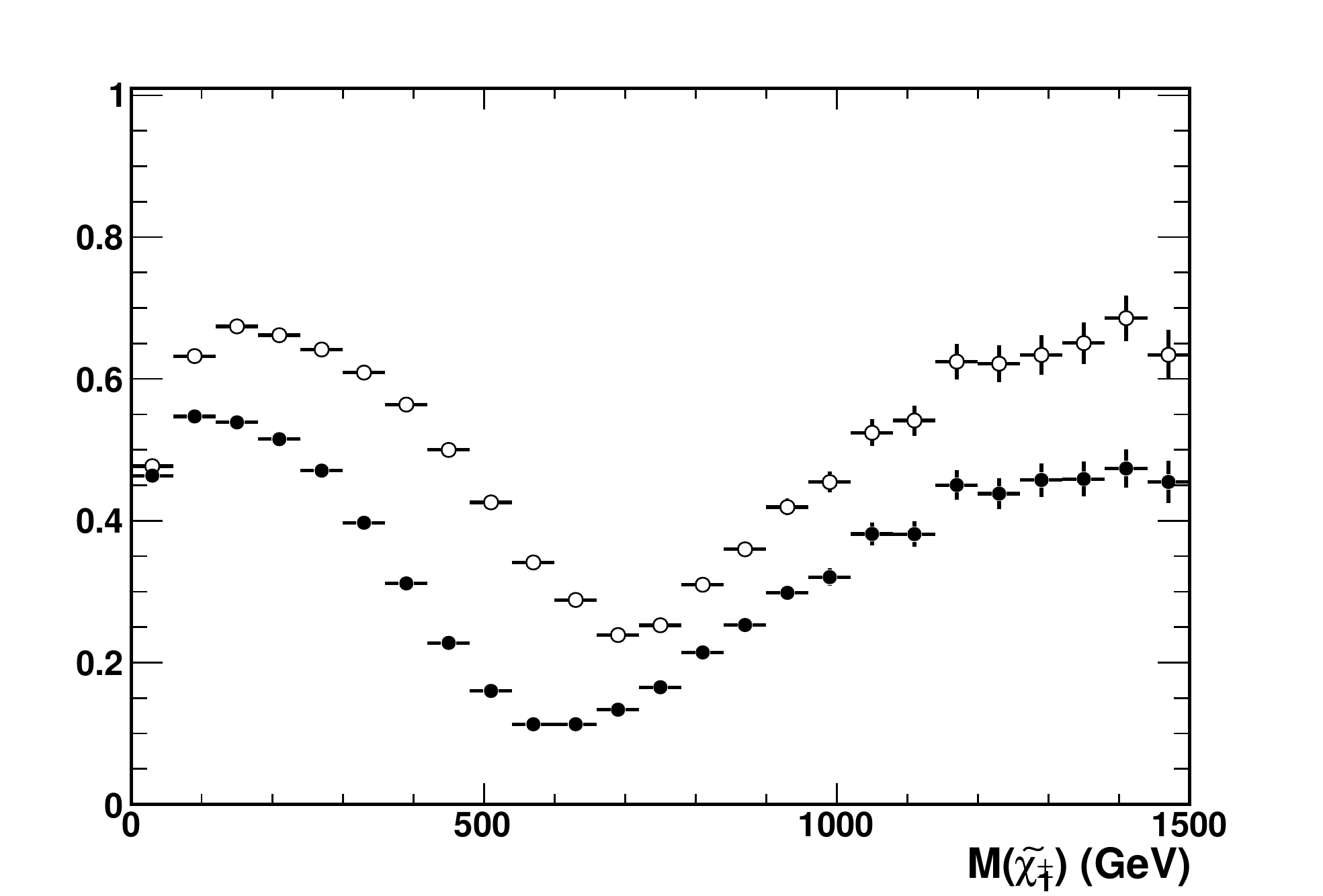} \\
\includegraphics[width=0.42\textwidth]{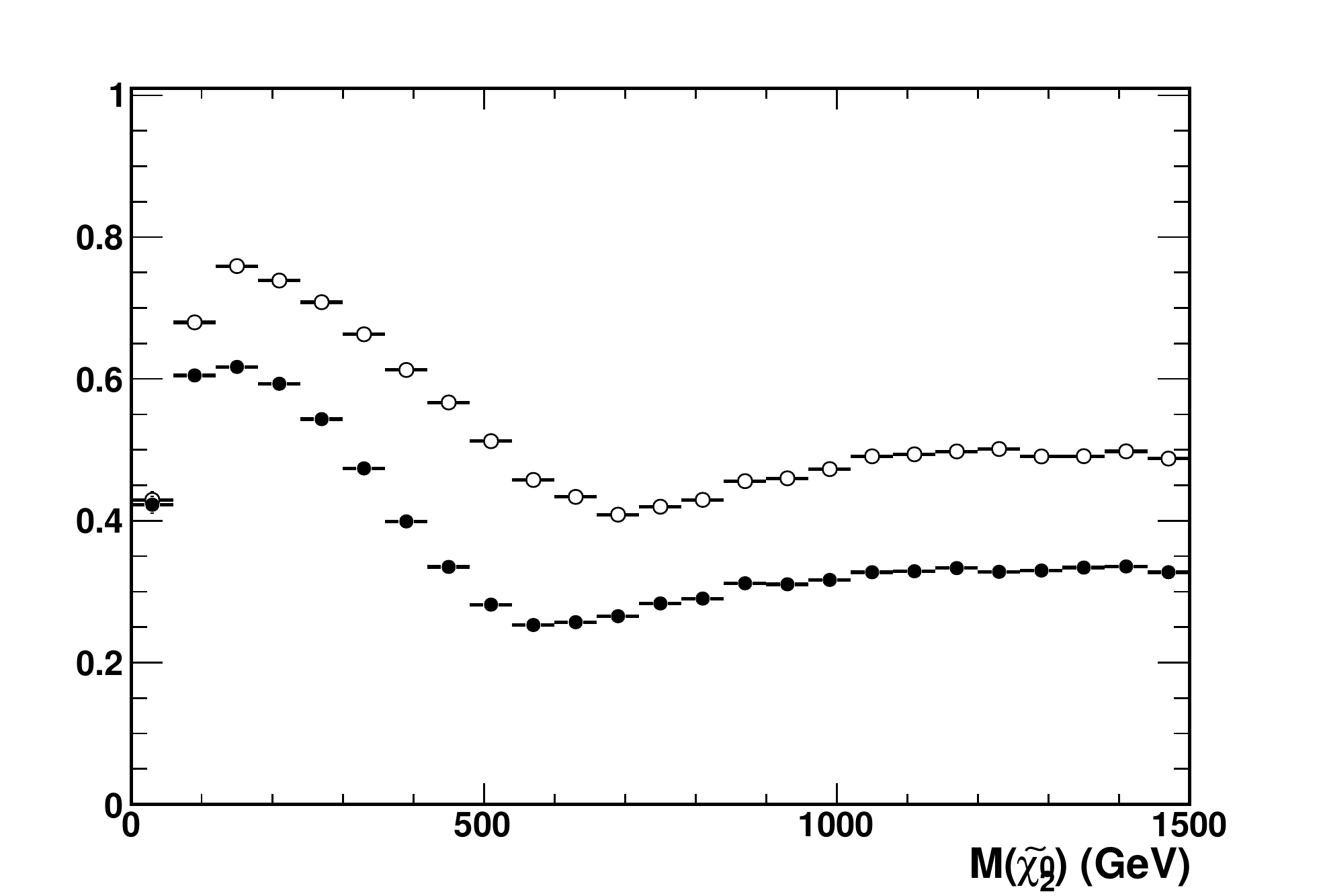}\includegraphics[width=0.42\textwidth]{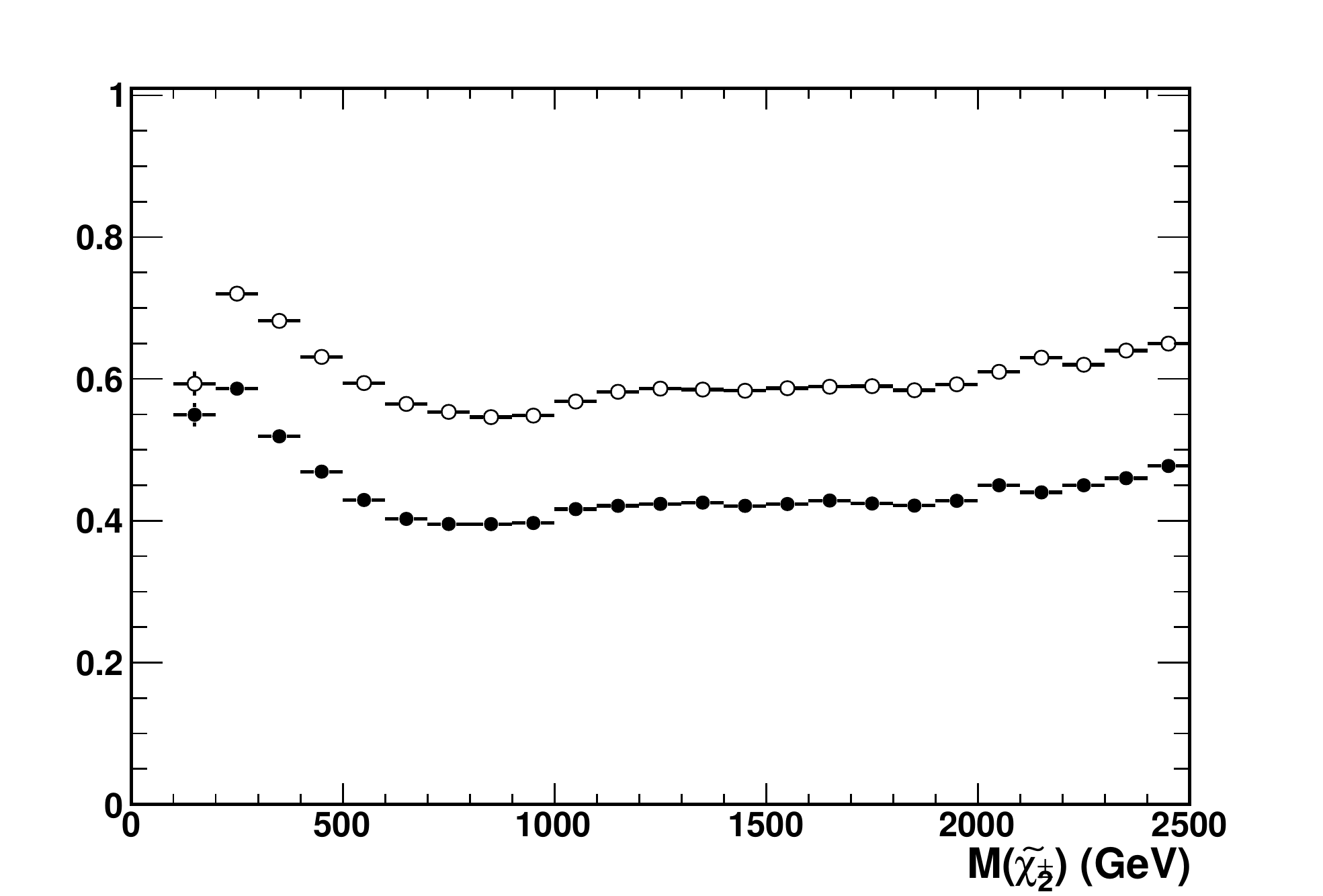} \\
\includegraphics[width=0.42\textwidth]{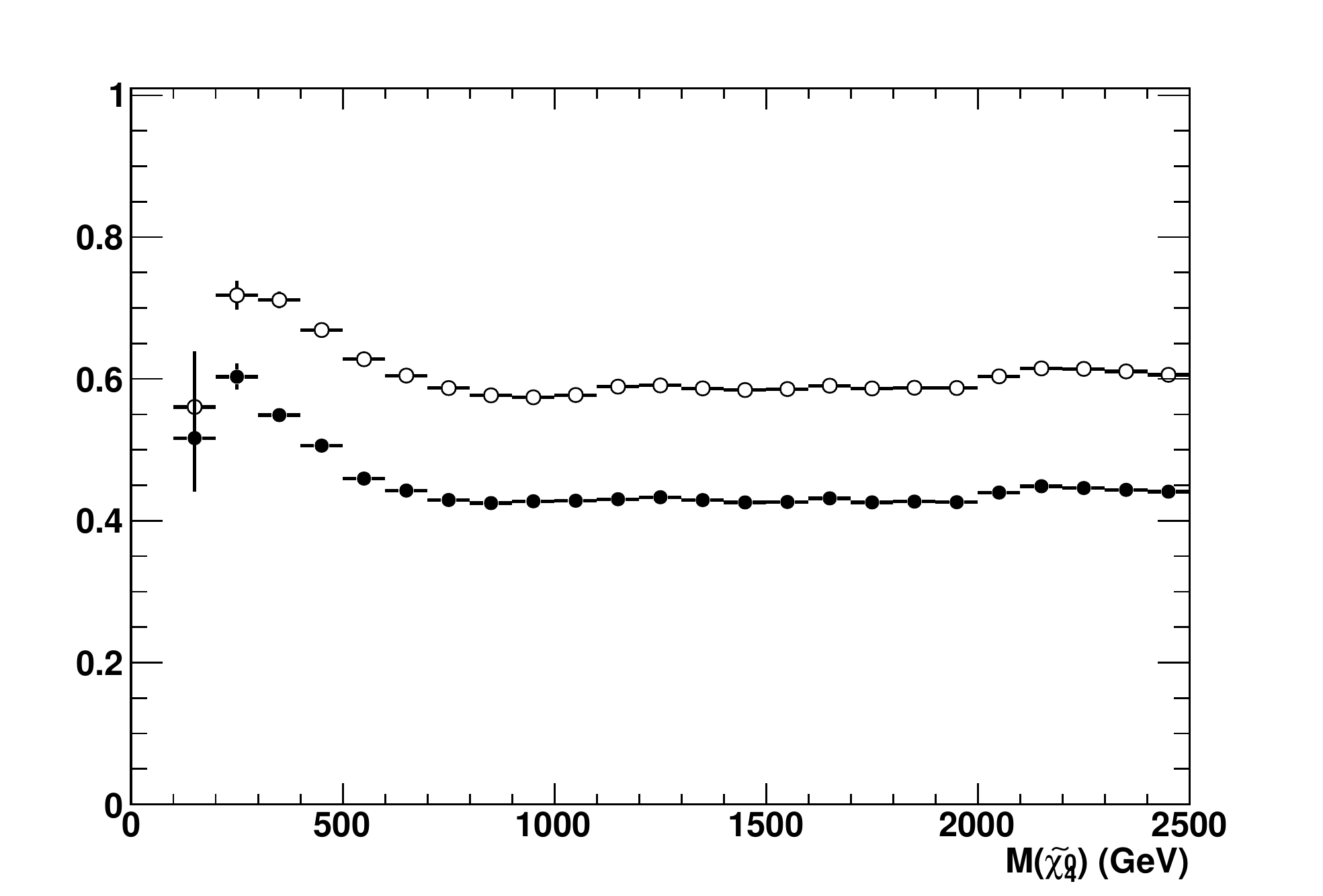}\includegraphics[width=0.42\textwidth]{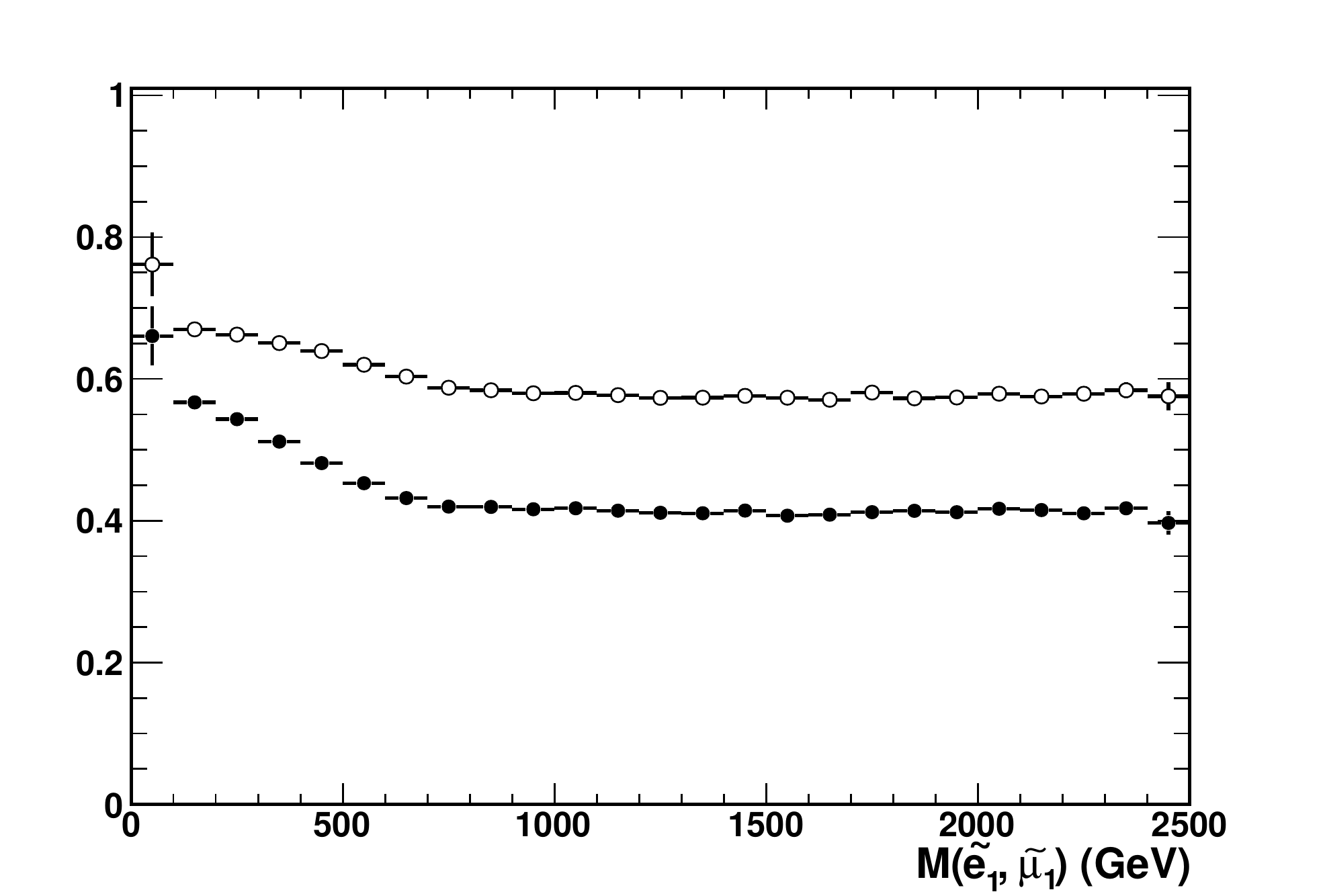} 
\end{tabular}
\end{center}
\caption{Fraction of valid MSSM points excluded at 95\% C.L. by 1~fb$^{-1}$ and 15~fb$^{-1}$ of 
LHC data as a function of the masses of $\tilde \chi^0_1$ (upper left), $\tilde \chi^{\pm}_1$ 
(upper right), $\tilde \chi^0_2$ (mid left), $\tilde \chi^{\pm}_2$ (mid right), 
$\tilde \chi^0_2$ (lower left) and the lightest slepton of the first two generations 
$\tilde \ell^{\pm}$ (lower right).}
\label{fig:w1fb}
\end{figure*}
Results are again presented as the fractions 
of accepted pMSSM points, which can be excluded by the results of the CMS 
analyses on 1~fb$^{-1}$ and 15~fb$^{-1}$ of data as a function of SUSY particle masses (see Figure~\ref{fig:w1fb}). As expected, the CMS analyses 
preferentially reduce the solutions with smaller masses for the sleptons 
and lighter gaugino states. However, $\sim$50\% of the accepted pMSSM 
points with masses below 400~GeV are not excluded within our scans. We also observe 
that the domain of SUSY weakly-interacting particle masses above 500~GeV 
is virtually unaffected by the present LHC data. This result contrasts the 
indications obtained in highly constrained SUSY models. These observations are 
important both for motivating searches of direct gaugino production at the LHC 
and for interpreting the LHC data to guide the choice of the energy scale of a 
future lepton collider.

Similarly to what done for strongly interacting sparticles, we assess the sensitivity of these results on the adopted range for the input SUSY parameters and the constraints. Here we study the change of the fractions of accepted pMSSM points, excluded with $M_{\tilde \chi^0_2}<$ 400~GeV and $M_{\tilde l}<$ 400~GeV on a dedicated scan. These fractions are 55.9\% and 57.1\% for our standard ranges with 1~fb$^{-1}$ of data. They become 37.8\% and 38.2\% if we increase the SUSY parameter range by a factor 1.5, 38.1\% and 37.9\% if we loosen the constraints (\ref{eq:bsg})-(\ref{eq:leptonic}) to a 3.5~$\sigma$ range, and 66.4\% and 68.9\% for a relic dark matter density cut corresponding to the 95\% C.L. of the seven year WMAP result. The large change with the range of SUSY parameters is expected since the LHC data has little or no sensitivity on the mass of these particles and the fractions change with the broadening of the parameter space.

\subsection{Dark matter direct detection}
\label{sec:4-3}

\begin{figure*}[t!]
\begin{center}
\begin{tabular}{c}
\includegraphics[width=0.49\textwidth]{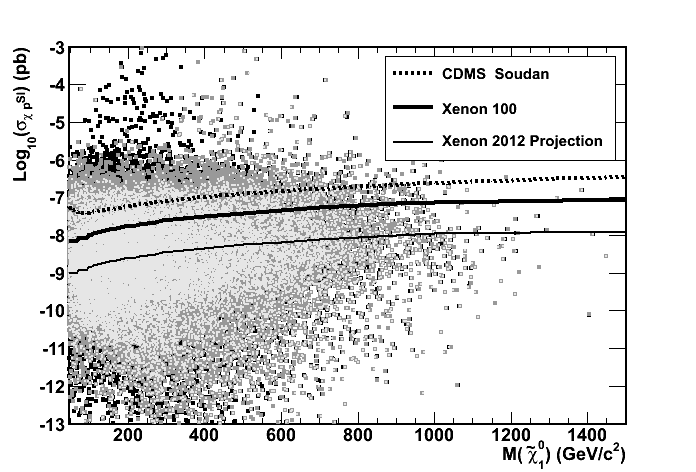} \\
\end{tabular}
\end{center}
\caption{$\tilde \chi p$ scattering cross section as a function of the LSP mass. The black 
dots represent accepted pMSSM points, the grey dots the subset of points not 
excluded by 1~fb$^{-1}$ and the light grey dots those not excluded by 15~fb$^{-1}$ of LHC data. The lines represent the 90\% C.L. exclusion contours set by the CDMS (dotted) and Xenon~100 (thick continuous) experiments together with the Xenon projection for the 2012 run (thin continuous).}
\label{fig:dd}
\end{figure*}

\begin{figure*}[ht!]
\begin{center}
\begin{tabular}{c}
\includegraphics[width=0.42\textwidth]{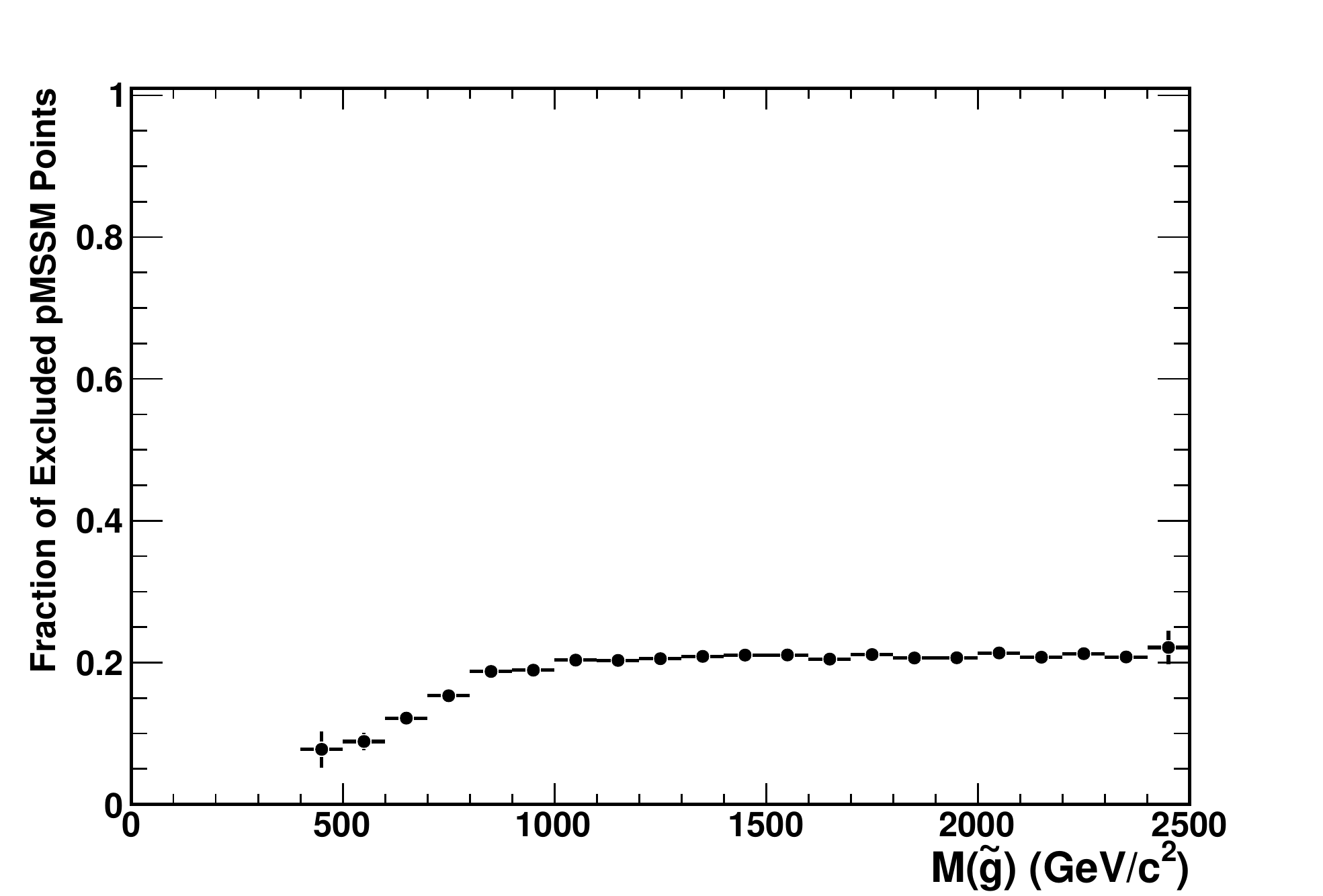}\includegraphics[width=0.42\textwidth]{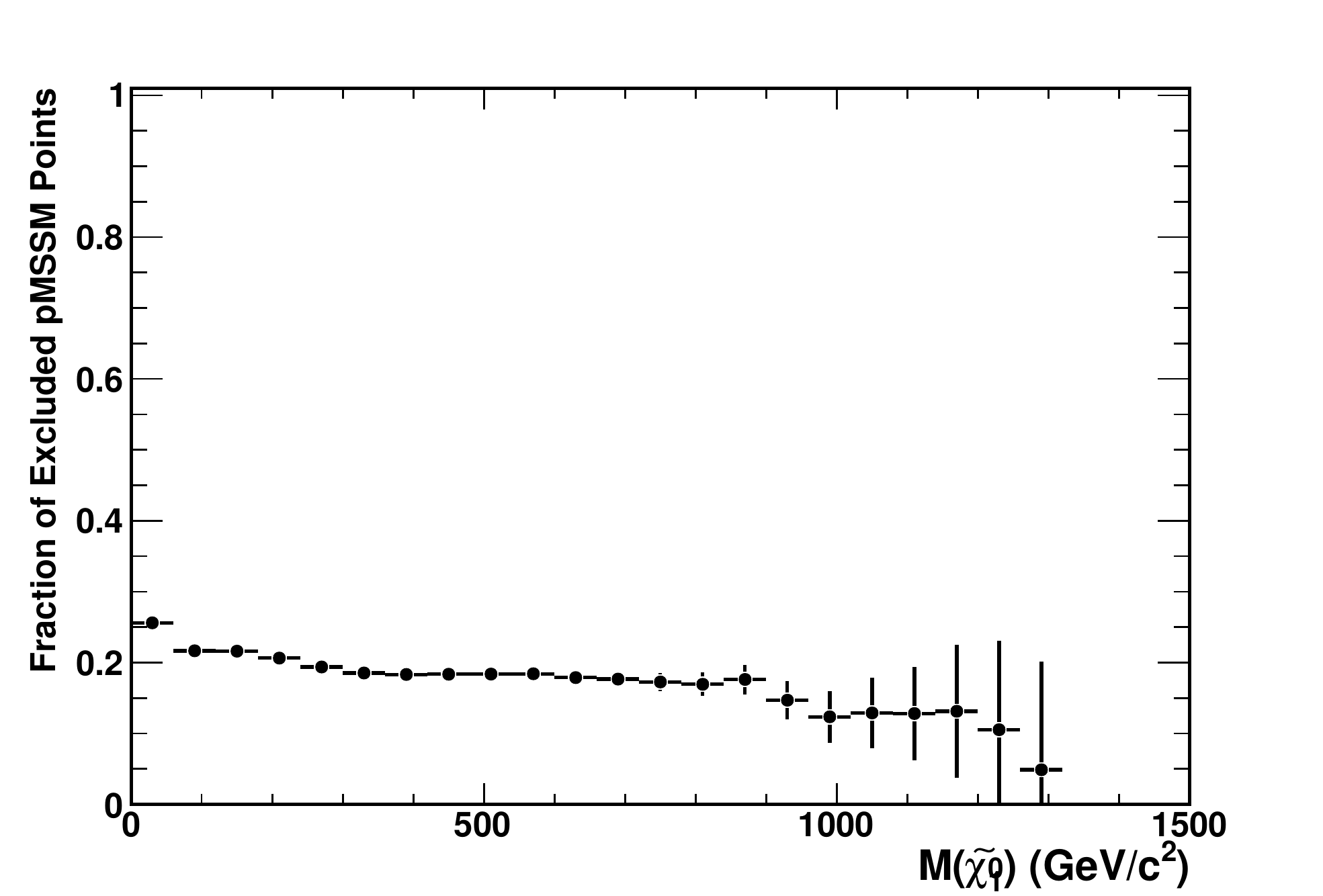} \\
\includegraphics[width=0.42\textwidth]{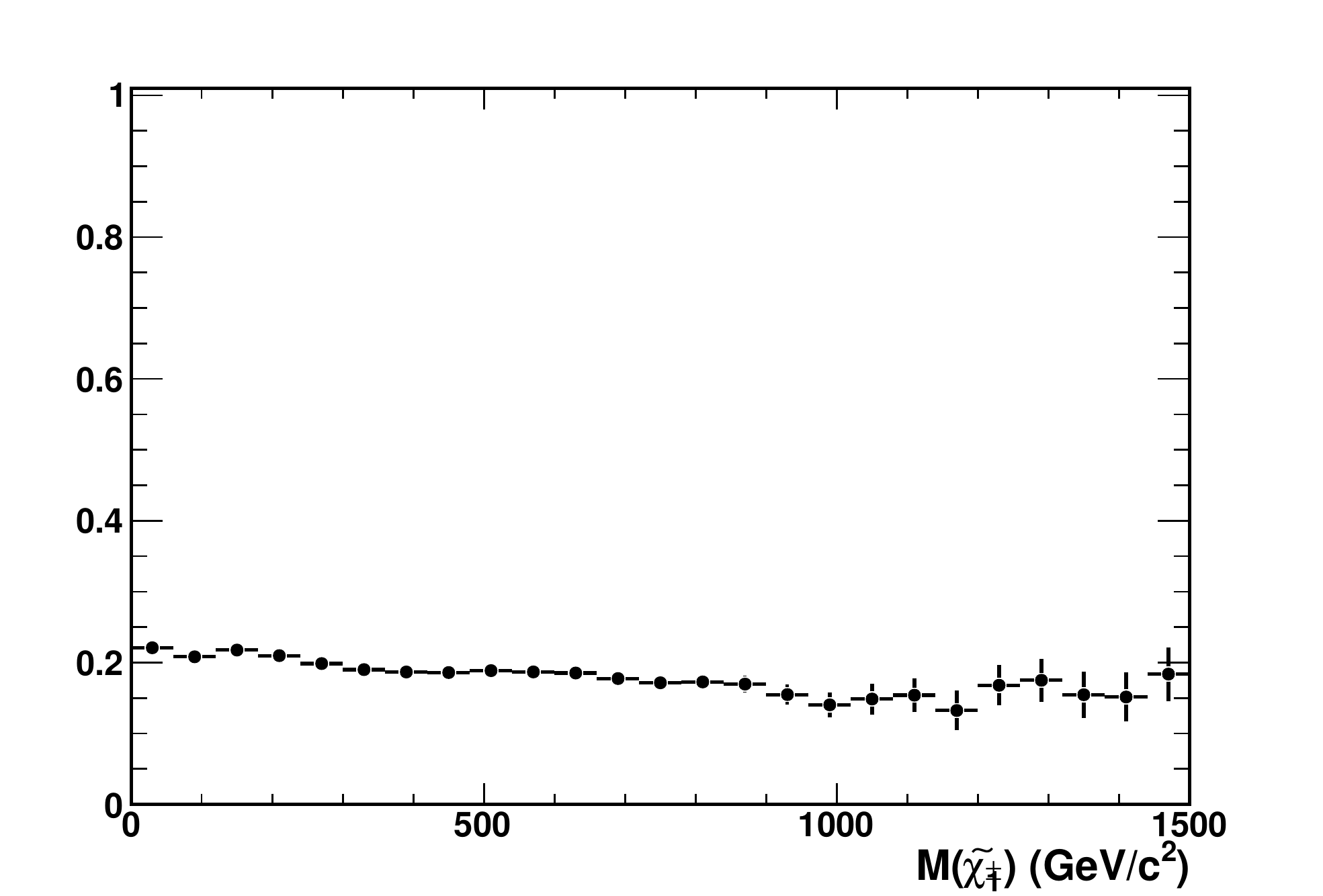}\includegraphics[width=0.42\textwidth]{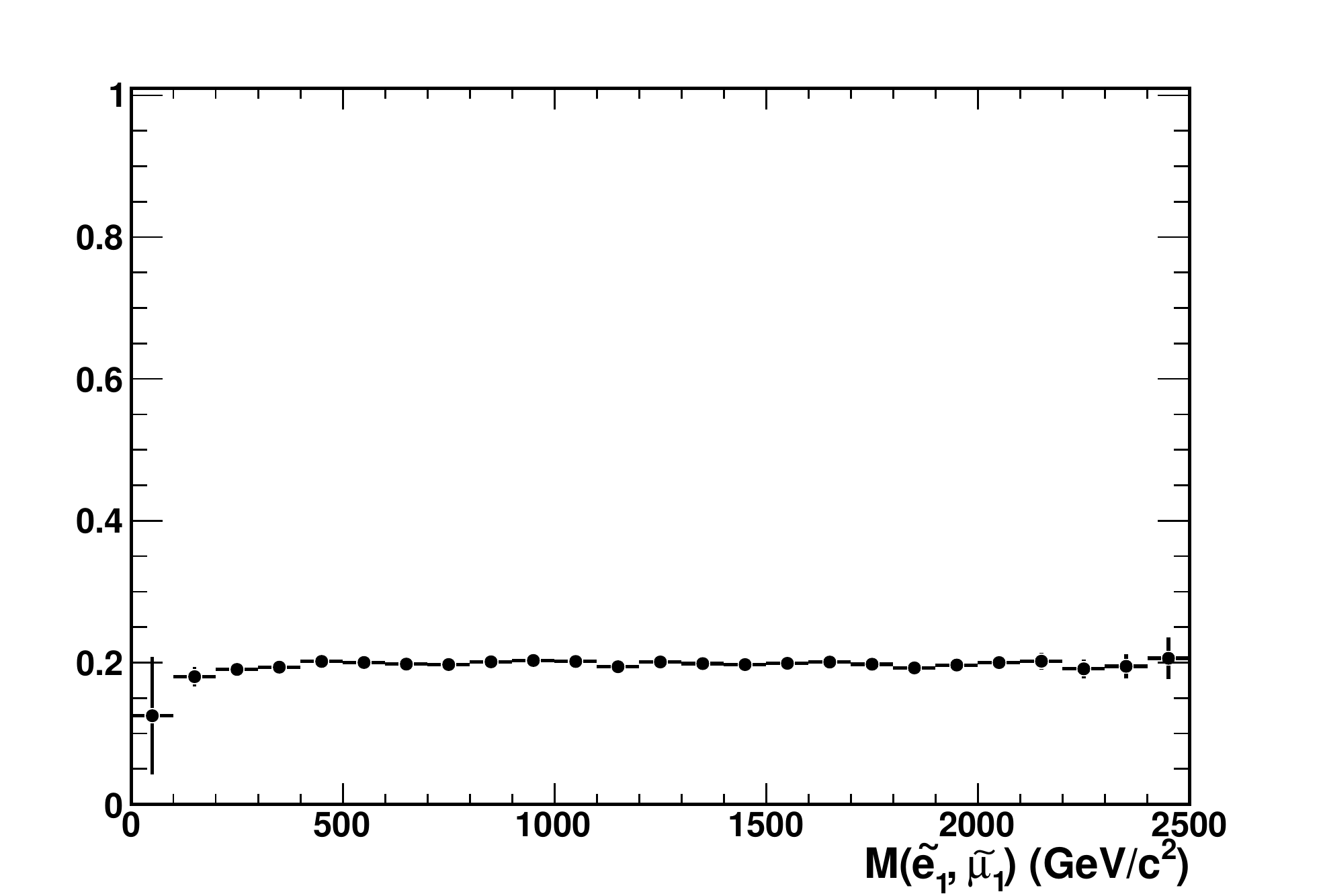}
\end{tabular}
\end{center}
\caption{Fraction of accepted MSSM points not excluded by 1~fb$^{-1}$ of LHC data but excluded at 90\% C.L. by the Xenon~100 direct detection data as a function of the masses of $\tilde g$ (upper left), $\tilde \chi^0_1$ (upper right), $\tilde \chi^{\pm}_1$ 
(lower left) and the lightest slepton of the first two generations $\tilde \ell^{\pm}$ (lower right).\vspace*{0.5cm}}
\label{fig:dd2}
\end{figure*}

The broad coverage of the pMSSM phase space through the scans provides us with 
an opportunity to contrast the impact of the LHC SUSY searches and the dark matter 
direct detection experiments. In particular, the recent results by the CDMS \cite{Ahmed:2009zw} and Xenon~100 \cite{Aprile:2011hi}
collaborations have moved the sensitivity of these experiments well within the 
region characteristics of the MSSM with $\tilde \chi^0_1$ LSP \cite{Farina:2011bh}. Figure~\ref{fig:dd} 
shows the distribution of the spin independent $\tilde \chi p$ scattering cross section as a function of 
the LSP mass for all the accepted pMSSM points and for those points not excluded by 
1 and 15~fb$^{-1}$ of LHC data together with the 90\%~C.L. exclusion bounds from CDMS and Xenon. The current 90\%~C.L. exclusion contour of the Xenon~100
experiment cuts through this region with about 20\% of the accepted pMSSM points 
giving a cross section exceeding the Xenon~100 limit. We note that 1~fb$^{-1}$ of LHC data preferentially removes points with light $\tilde \chi^0_1$ in the large $\tilde \chi p$ scattering cross section region which is also incompatible with direct detection experiment data.
Since the main contribution to the $\tilde \chi p$ cross section comes from diagrams mediated by the $A^0$ boson, the neutralino dark matter direct detection exclusion region preferentially removes 
points at low $M_A$ and large $\tan \beta$ values. However, this constraint does not 
significantly impact the mass distributions for other SUSY particles. The fraction 
of accepted pMSSM points incompatible with the Xenon~100 data is shown as a function 
of the mass of the gluino, lightest slepton of the first two generations,
$\tilde \chi^{\pm}_1$  and $\tilde \chi^{0}_2$ in Figure~\ref{fig:dd2}.

\subsection{Higgs couplings}
\label{sec:4-4}

The search for the Higgs boson is one of the most important themes of the present LHC physics program.
The current data has narrowed the allowed region for a SM-like Higgs boson down to mass interval 
114~GeV $< M_H <$ 145~GeV. This is a tantalising indication that the likely mass region 
for the Higgs boson exactly coincides with that favoured by electroweak precision data for a SM Higgs boson 
and by SUSY for the lightest Higgs state, $h^0$. The points in our scan which are not excluded by 1~fb$^{-1}$ of LHC data have $M_H$ in the interval 98 to 130~GeV. The average value of $M_H$ for these points is 118.2 GeV and we do not observe any significant shift compared to the value for all the accepted points. By the end of the 2012 the LHC should deliver enough luminosity, 
for both ATLAS and CMS to exclude the existence of a SM-like Higgs boson, if no signal is observed. 
It is important to understand how this prediction translates for the case of $h^0$. 
Its couplings to fermions and gauge bosons are shifted by SUSY corrections \cite{Djouadi:2005gj}. 
As a result, in SUSY both the production cross section $gg \to h^0$ and the branching fractions for the decay 
processes $h^0 \to \gamma \gamma$ and $W^+ W^-$ 
differ from their SM values. We quantitatively study these differences for the accepted pMSSM 
points which are not excluded by the CMS searches on 1~fb$^{-1}$ and 15~fb$^{-1}$ of statistics, as discussed above. 
The single most sensitive variable is the mass of the $A^0$ boson.

\begin{figure}[t!]
\begin{center}
\begin{tabular}{c}
\includegraphics[width=0.42\textwidth]{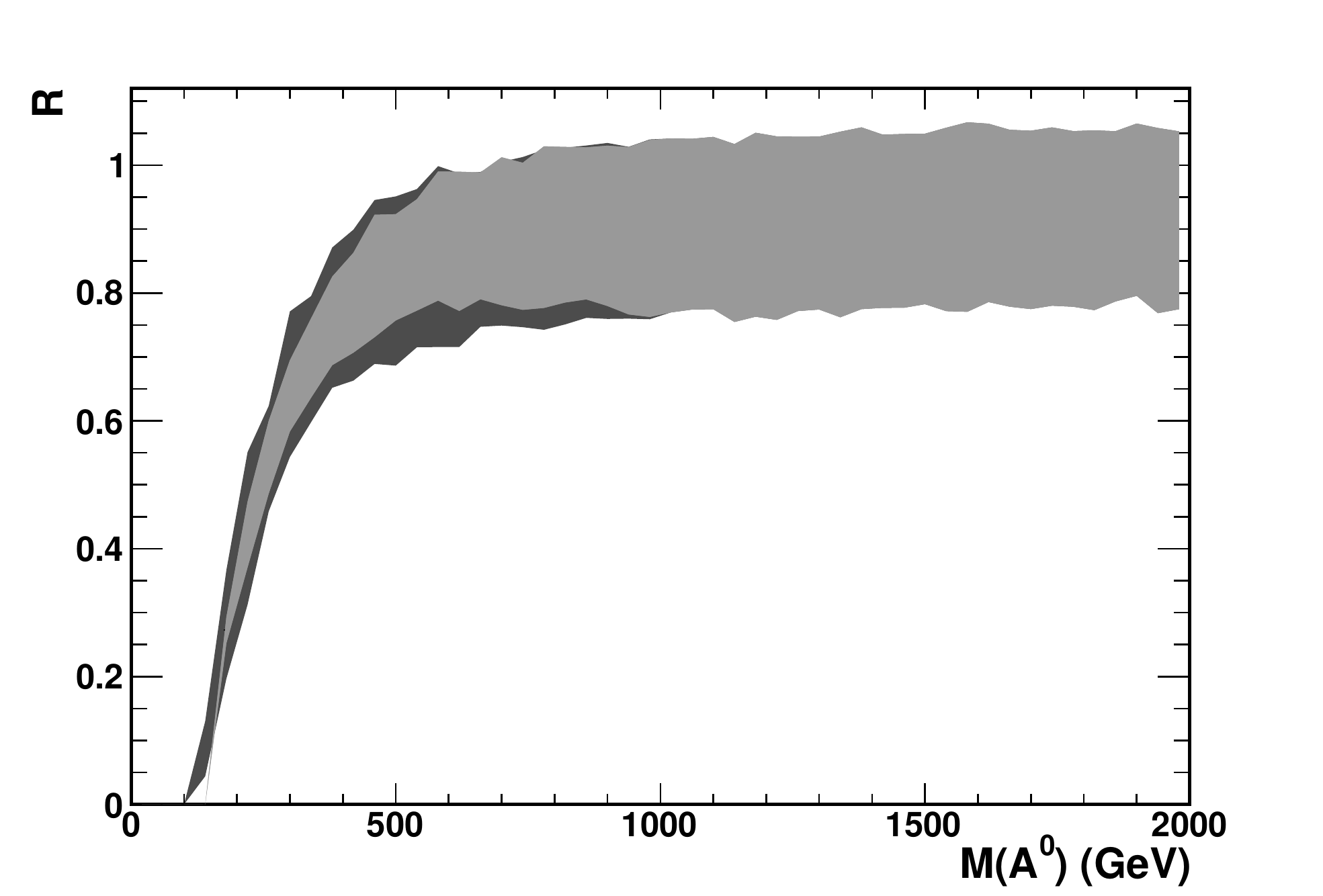} \\
\end{tabular}
\end{center}
\caption{Distribution of the ratio $R = \frac{\sigma (gg \to h^0)}{\sigma (gg \to H^0_{SM} )} 
\times \frac{{\mathrm{BR}}(h^0 \to \gamma \gamma)}{{\mathrm{BR}}(H^0_{SM} \to \gamma \gamma)}$ as 
a function of $M_A$ for accepted pMSSM points not excluded by 1~fb$^{-1}$ (dark grey) and 15~fb$^{-1}$ (light grey) of LHC data. The shaded area shows the full range of the values of $R$ over the accepted pMSSM points.}
\label{fig:hratio}
\end{figure}

\begin{figure}[t!]
\begin{center}
\begin{tabular}{c}
\includegraphics[width=0.42\textwidth]{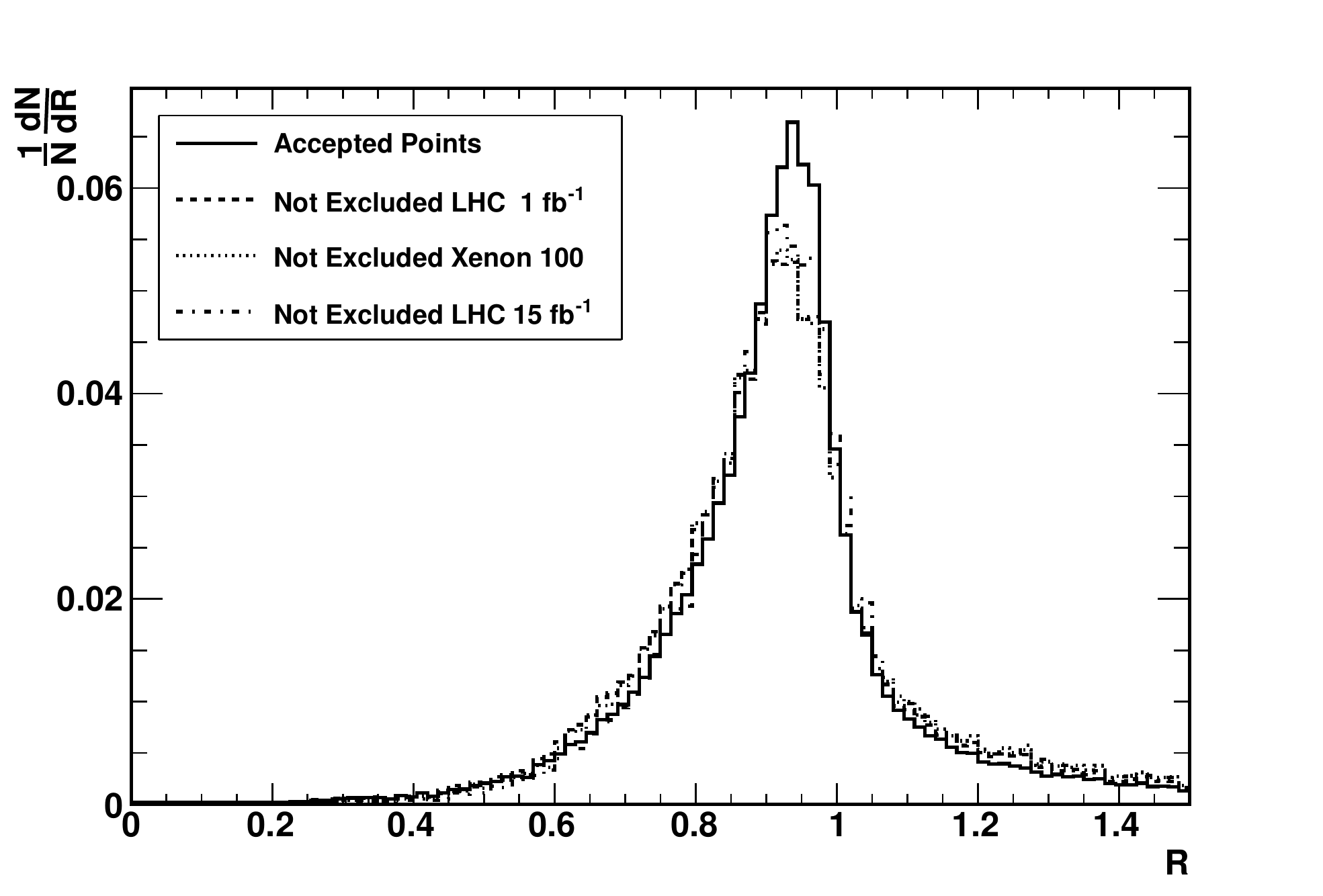} \\
\end{tabular}
\end{center}
\caption{Distribution of the ratio $R = \frac{\sigma (gg \to h^0)}{\sigma (gg \to H^0_{SM} )} 
\times \frac{{\mathrm{BR}}(h^0 \to \gamma \gamma)}{{\mathrm{BR}}(H^0_{SM} \to \gamma \gamma)}$ for accepted pMSSM points (solid line), the same and not excluded by 1~fb$^{-1}$ of LHC data (dashed line), the same and not excluded by Xenon~100 data (dotted line) and the same and not excluded by 15~fb$^{-1}$ of LHC data (dashed-dotted line).\vspace*{1.cm}}
\label{fig:hratio2}
\end{figure}

Figure~\ref{fig:hratio} shows the range of values taken by the ratio 
$R = \frac{\sigma (gg \to h^0)}{\sigma (gg \to H^0_{SM} )} 
\times \frac{{\mathrm{BR}}(h^0 \to \gamma \gamma)}{{\mathrm{BR}}(H^0_{SM} \to \gamma \gamma)}$ as a function 
of $M_A$ for accepted pMSSM points and not excluded 
by the CMS searches for a heavy Higgs boson and for SUSY particles with 1~fb$^{-1}$ and 15~fb$^{-1}$ of data as a function of $M_A$. The LHC searches for 
strongly interacting particles do not significantly change the prediction of the SUSY suppression for the product of production cross section and decay branching fraction, $R$. Figure~\ref{fig:hratio2} shows the distribution of $R$ for all the accepted pMSSM points, for the points not excluded by the LHC searches for gluinos and scalar quarks and also for the points compatible with the Xenon~100 direct detection experiment. The 2011 + 2012 statistics will enable tighter limits on the combination of $M_A$ and $\tan \beta$, if no signal is observed. However, it is interesting to observe that pMSSM points with suppression of the product of production cross section 
and branching fraction compared to the SM value of a factor of two or larger will remain. Even in the decoupling regime, \textit{i.e.}\ for $M_A >$ 500~GeV and beyond the $A^0$ boson sensitivity of the LHC 7~TeV run, there are acceptable pMSSM solutions where $R$ is as small as 0.75. 


\subsection{Fine tuning}
\label{sec:4-5}

A quantitative way of measuring how natural a model point is, can be achieved by introducing the fine tuning parameter. We use here an analogous to the fine tuning measure introduced first in \cite{Ellis:1986yg} and \cite{Barbieri:1987fn}, although we follow mostly the approach presented in \cite{Perelstein:2007nx} and focus on the amount of fine tuning at the weak scale, rather than at the GUT scale, of the MSSM parameters.

The mass of the $Z$ boson in the MSSM is given by

\begin{equation}
M_Z^2 \,=\, - M_{H_u}^2 \left( 1-\frac{1}{\cos 2 \beta} \right) - 
M_{H_d}^2 \left( 1+\frac{1}{\cos 2 \beta} \right)-2|\mu|^2
\end{equation}
at tree level, where 
\begin{equation}
\sin 2 \beta  = \frac{2 b}{M_{H_u}^2 + M_{H_d}^2+2|\mu|^2}\,.
\end{equation}
Following \cite{Ellis:1986yg,Barbieri:1987fn}, we quantify fine tuning by computing
\begin{equation}
\delta(\xi)\,=\,\left| \frac{\partial\log M_Z^2}{\partial\log \xi}\right|,
\end{equation}
where $\xi=M_{H_u}^2, M_{H_d}^2, b$ and $\mu$ are the relevant parameters. This gives:

\begin{eqnarray}
\delta(\mu) &=& \frac{4\mu^2}{M_Z^2}\,\left(1+\frac{M_A^2+M_Z^2}{M_A^2}
\tan^2 2\beta \right), \nonumber \\ 
\delta(b) &=& \left( 1+\frac{M_A^2}{M_Z^2}\right)\tan^2 2\beta, \nonumber \\
\delta(M_{H_u}^2) &=& \left| \frac{1}{2}\cos2\beta +\frac{M_A^2}{M_Z^2}\cos^2\beta
-\frac{\mu^2}{M_Z^2}\right|\nonumber \\
&& \times\left(1-\frac{1}{\cos2\beta}+
\frac{M_A^2+M_Z^2}{M_A^2} \tan^2 2\beta \right), \nonumber \\ 
\delta(M_{H_d}^2) &=& \left| -\frac{1}{2}\cos2\beta +\frac{M_A^2}{M_Z^2}\sin^2\beta
-\frac{\mu^2}{M_Z^2}\right|\nonumber \\
&& \times\left|1+\frac{1}{\cos2\beta}+
\frac{M_A^2+M_Z^2}{M_A^2} \tan^2 2\beta \right|. \nonumber
\end{eqnarray}

\begin{figure}[t!]
\begin{center}
\begin{tabular}{c}
\includegraphics[width=0.42\textwidth]{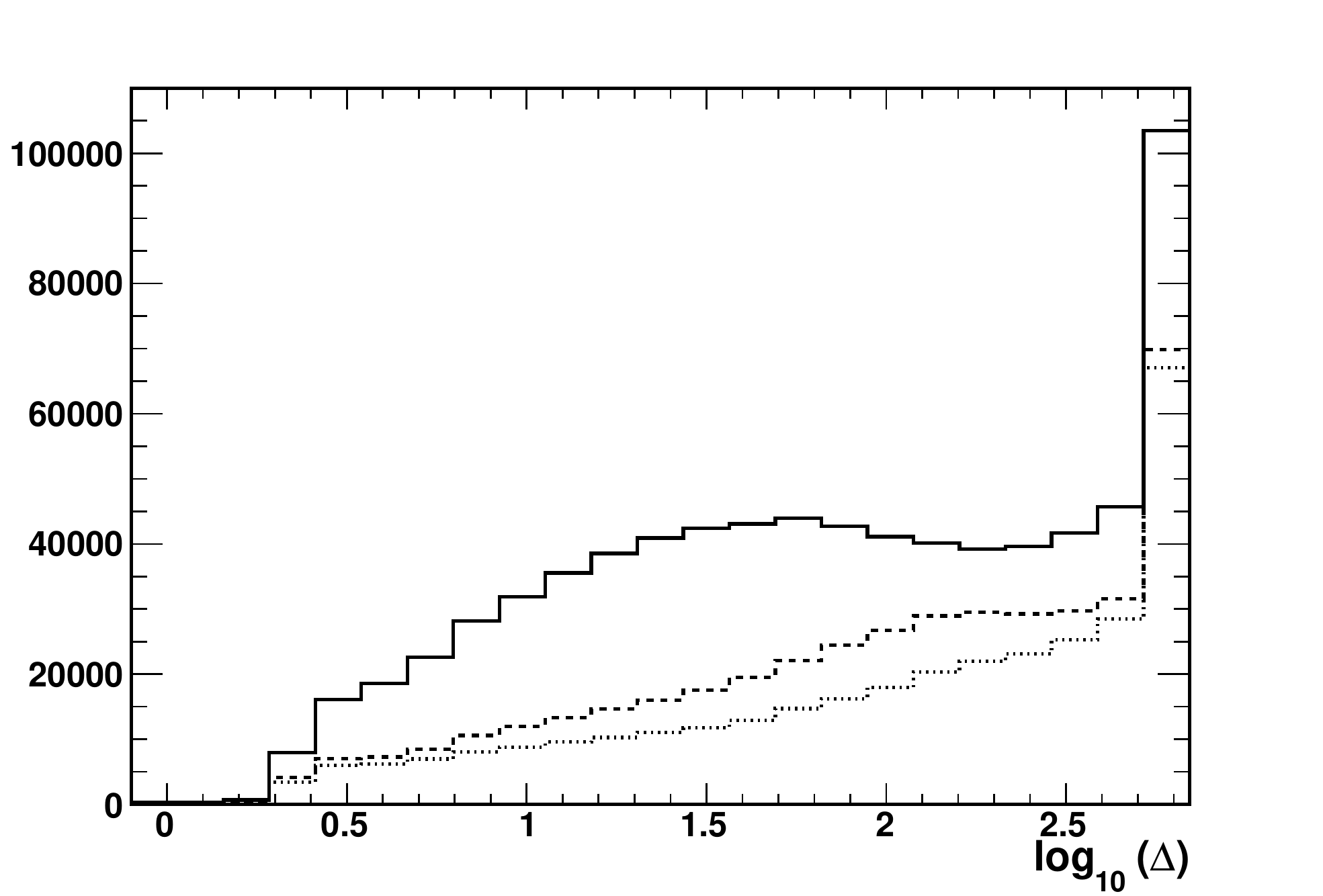} 
\end{tabular}
\end{center}
\caption{Distributions of the fine tuning variable $\Delta$ after the flavour, $g_\mu-2$ and $\Omega_{\mathrm{DM}} h^2$ (solid line), LHC (dashed line) and Xenon~100 (dotted line) subsequent constraints. The last bin includes overflows.}
\label{fig:ft}
\end{figure}

The overall fine tuning is obtained by adding the four $\delta$'s in quadrature: 
\begin{equation}
\Delta=\big[\delta(\mu)^2+ \delta(b)^2+\delta(M_{H_u}^2)^2+\delta(M_{H_d}^2)^2\big]^{1/2}\;.
\end{equation}
The larger this value, the more fine tuned is the SUSY parameter point. As a reference, the well known SPS1a point \cite{Allanach:2002nj} (now excluded by LHC data) has a fine tuning of about 70 in this definition. Figure~\ref{fig:ft} shows the distribution of fine tuning of our accepted pMSSM points, 
of those not excluded by the LHC data and by the Xenon~100
scattering cross section upper bound. 
The shape of the fine tuning distribution is modified by these constraints, which preferentially removes particles with low masses, shifting the distribution towards larger fine tuning values.
However, a significant population of pMSSM solutions at low to moderate values of fine tuning survives after these cuts.

\section{Conclusions}
\label{sec:5}
The implications of LHC searches on SUSY particle spectra have been studied using flat scans of the 19-parameters pMSSM phase space. We evaluate the sensitivity of flavour physics observables to SUSY particle masses, and apply them as constraints together with $g_\mu-2$, dark matter and the mass bounds from earlier LEP and Tevatron searches. The sensitivity of the LHC SUSY searches with jets, leptons and missing energy is assessed by reproducing the recent CMS analyses with fast simulation, validated on benchmark points. We study the fraction of pMSSM points compatible with all the constraints which are excluded by the LHC searches with 1 and 15~fb$^{-1}$ as a function of the mass of strongly and weakly interacting SUSY particles on a sample of 835k accepted points, without performing an importance weighting of the points. We observe that the LHC data can exclude more than 85\% of these pMSSM points up to a gluino mass of $\sim$520~GeV 
and 700~GeV for 1~fb$^{-1}$ and 15~fb$^{-1}$, respectively. On the contrary, the mass spectra of most of colour singlet states are only weakly impacted and, for the range of SUSY parameters adopted in this study, $\sim$50\% of the accepted points with masses below 400~GeV are not excluded. In the pMSSM, the domain of SUSY weakly-interacting particle masses above 500~GeV is virtually unaffected by the present LHC data contrary to the results in highly constrained scenarios. 
Compared to the study of \cite{Sekmen:2011cz}, which is based on a similar methodology for assessing the SUSY sensitivity of the LHC with CMS analyses, the results reported here for the impact of 1~fb$^{-1}$ of LHC data are in agreement.

Comparing to dark matter direct detection experiments, the 95\%~C.L. exclusion contour of the Xenon~100
experiment cuts through the region of the accepted pMSSM points with about 20\% of them exceeding the Xenon~100 bound. The LHC data preferentially removes points in the large $\tilde \chi p$ scattering cross section. A significant fraction of pMSSM solutions compatible with both the LHC and dark matter direct detection experiment data have low to moderate values of the fine tuning parameter. 
Finally, we use our sample of accepted pMSSM points to estimate the suppression of the light Higgs production cross section and $\gamma\gamma$ decay branching fractions compared to the Standard Model predictions. We find that a suppression of up to more than a factor of two is compatible both with the current LHC data and with that to be collected by the end of 2012, which has implications on the Higgs boson detectability.

\section*{Acknowledgements}
We would like to thank M.~Mangano for supporting this activity and the LPCC for making dedicated  computing resources available to us. Several colleagues in the CMS collaboration provided us with information and feedback, in particular A.~De Roeck, O.~Buchmueller, B.~Hooberman and M.~Bryn. We acknowledge useful discussions with J.~Hewett, T.~Rizzo, S.~Kraml and S.~Heinemeyer. We are grateful to P.~Skands for help with PYTHIA, B.~Allanach with Softsusy, T.~Stefaniak and K.~Williams with HiggsBounds, and E.~Gianolio for computing support. E.~Aprile, P.~Beltrame and A.~Melgarejo provided us with information on the XENON results, B. Sadoulet and D.~Speller with CDMS, which we gratefully acknowledge. Moreover, we specially thank S.~Sekmen for help with event simulation. 

%
\bibliographystyle{epjc}
\bibliography{susyscan1.bib}

\end{document}